\documentclass[aps,prm,twocolumn,floatfix,superscriptaddress]{revtex4}
\usepackage{times}
\usepackage{graphicx}
\usepackage{epsfig}
\usepackage{amsmath}
\usepackage{amssymb}
\usepackage[version=3]{mhchem}
\usepackage{xcolor}
\usepackage{multirow}

\begin{document}

\title{Structure and dynamics of the negative thermal expansion material Cd(CN)$_{\boldsymbol2}$\\ under {\color{black} hydrostatic} pressure}

\author{Chloe S. Coates}
\affiliation{Department of Chemistry, University of Oxford, Inorganic Chemistry Laboratory, South Parks Road, Oxford OX1 3QR, U.K.}

\author{Mia Baise}
\affiliation{Department of Chemistry, 20 Gordon Street,  University College London, WC1H 0AJ, U.K.}

\author{Johnathan M. Bulled}
\affiliation{Department of Chemistry, University of Oxford, Inorganic Chemistry Laboratory, South Parks Road, Oxford OX1 3QR, U.K.}
\affiliation{Department of Materials, ETH Zürich, Vladimir-Prelog-Weg 4, 8093, Zürich, Switzerland.}

\author{Emma H. Wolpert}
\affiliation{Department of Chemistry, University of Oxford, Inorganic Chemistry Laboratory, South Parks Road, Oxford OX1 3QR, U.K.}
\affiliation{Department of Chemistry, 20 Gordon Street,  University College London, WC1H 0AJ, U.K.}

\author{Joshua W. Makepeace}
\affiliation{Department of Chemistry, University of Oxford, Inorganic Chemistry Laboratory, South Parks Road, Oxford OX1 3QR, U.K.}
\affiliation{School of Chemistry, University of Birmingham, Edgbaston, B15 2TT, U.K.}

\author{Helen C. Walker}
\affiliation{ISIS Facility, Rutherford Appleton Laboratory, STFC, Didcot, OX11 0QX, U.K.}

\author{Alexandra S. Gibbs}
\affiliation{ISIS Facility, Rutherford Appleton Laboratory, STFC, Didcot, OX11 0QX, U.K.}
\affiliation{School of Chemistry, University of St Andrews, North Haugh, St Andrews KY16 9ST, U.K.}

\author{A. Dominic Fortes}
\affiliation{ISIS Facility, Rutherford Appleton Laboratory, STFC, Didcot, OX11 0QX, U.K.}

\author{Ben Slater}
\affiliation{Department of Chemistry, 20 Gordon Street,  University College London, WC1H 0AJ, U.K.}

\author{Andrew L. Goodwin}
\email{andrew.goodwin@chem.ox.ac.uk}
\affiliation{Department of Chemistry, University of Oxford, Inorganic Chemistry Laboratory, South Parks Road, Oxford OX1 3QR, U.K.}

\date{\today}
\begin{abstract}
 We use a combination of variable-temperature / variable-pressure neutron powder diffraction, variable-pressure inelastic neutron scattering, and quantum chemical calculations to interrogate the behaviour of the negative thermal expansion (NTE) material $^{114}$Cd(CN)$_2$ under hydrostatic pressure. We determine the equation of state of the ambient-pressure phase, and discover the so-called `{\color{black}thermal} hardening' effect whereby the material becomes elastically stiffer as it is heated. We also identify a number of high-pressure phases, and map out the phase behaviour of Cd(CN)$_2$ over the range $0\leq p\leq0.5$\,GPa, $100\leq T\leq300$\,K. As expected for an NTE material, the low-energy phonon frequencies are found to soften under pressure, and we determine an effective Gr{\"u}neisen parameter for these modes. Finally, we show that the elastic behaviour of Cd(CN)$_2$ is sensitive to the local Cd coordination environment, which suggests an interplay between short- (phononic) and long-timescale (cyanide flips) fluctuations in Cd(CN)$_2$.
 \end{abstract}


\maketitle

\section{Introduction}

Negative thermal expansion (NTE) materials are useful and interesting because their structures contract---rather than expand---on heating \cite{Sleight_1998,Evans_1999,Barrera_2005,Takenaka_2012,Lind_2012,Chen_2015,Dove_2016}. Of the many different types of NTE systems now known, cadmium cyanide (Cd(CN)$_2$) is remarkable for a number of reasons. First, its NTE is isotropic: the Cd(CN)$_2$ structure has cubic symmetry \cite{Shugam_1945,Goodwin_2005,Fairbank_2012}, which is an attractive design consideration to minimise {\color{black} microcracking due to thermally induced stresses} \cite{Sleight_1998}. Second, the magnitude of its NTE effect is particularly large \cite{Goodwin_2005,Coates_2021b}---indeed many times greater than that of well-known isotropic NTE materials such ZrW$_2$O$_8$ \cite{Mary_1996} and ScF$_3$ \cite{Greve_2010}. And, third, this NTE persists over an extremely large temperature window of about 600\,K \cite{Coates_2021b}. In fact its combination of NTE strength and persistence gives it the {\color{black}one of the} largest NTE {\color{black}capacities} of all known materials \cite{Coates_2019,Coates_2021b,Takenaka_2024}.

Mechanistically it {\color{black}was originally} assumed that NTE arises in Cd(CN)$_2$ by the same processes operating in its better-studied congener Zn(CN)$_2$: namely, the thermal population of low-frequency phonons that act to draw the structure in on itself (so-called negative Gr{\"u}neisen modes) \cite{Goodwin_2005,Zwanziger_2007,Ding_2008}. But we {\color{black}have since discovered} that the cyanide ions in Cd(CN)$_2$ can in fact `flip', and these long-timescale excitations act to enhance NTE further \cite{Coates_2021}. There turns out to be a surprising parallel between CN$^-$ flips in Cd(CN)$_2$ and spin-flips in the dipolar spin-ices such as Dy$_2$Ti$_2$O$_7$ \cite{Bramwell_1998,denHertog_2000,Fennell_2009,Tomasello_2015}; this is a point to which we will return in due course.

Whatever the precise NTE mechanism may be, pressure is a particularly important thermodynamic variable for studying {\color{black}any} phonon-driven NTE {\color{black}system} \cite{Perottoni_1998,Jorgensen_1999,Chapman_2009,Poswal_2009,Greve_2010,Fang_2013,Ma_2025}. The Gr{\"u}neisen parameter quantifies the effect of volume change on phonon frequencies, and the negative values associated with NTE modes imply that the corresponding mode frequencies are unstable under pressure \cite{Mittal_2009}. Consequently, one expects to see displacive phase transitions in high-pressure studies of any phononic NTE material, with the distortions involved providing insight into the dominant NTE modes of the ambient phase \cite{Jorgensen_1999,Greve_2010,Collings_2013}. It is common too for isotropic NTE materials to exhibit other counterintuitive mechanical phenomena at high pressure, such as amorphisation \cite{Perottoni_1998,Secco_2001,Keen_2007,Hu_2010} or pressure-induced softening---whereby a material becomes softer, rather than harder, as it is compressed \cite{Fang_2013c}.

Consequently, we sought to explore the structural and dynamical behaviour of Cd(CN)$_2$ as a function of hydrostatic pressure. Our approach is to use a combination of variable-pressure neutron scattering measurements (both elastic and inelastic) and quantum chemical calculations. An extreme sensitivity to damage by X-rays \cite{Coates_2021b} rules out the use of variable-pressure X-ray diffraction measurements, and we highlight that our use of neutron scattering is feasible only because we have access to a $^{114}$Cd-enriched sample \cite{Coates_2018,Coates_2021}. Our key results are (i)  the measurement of the equation of state of ambient-phase Cd(CN)$_2$, (ii) the discovery of strong pressure-induced softening and `{\color{black}thermal} hardening' effects, (iii) explicit measurement of the pressure-induced phonon mode softening associated with negative Gr{\"u}neisen modes, (iv) the identification of a series of high-pressure phases and their thermal stability, and (v) the identification of interplay between cadmium coordination environments and elastic moduli.

Our paper begins with a brief review of what is known regarding NTE in Cd(CN)$_2$, and also some of the key implications of the spin-ice model developed elsewhere to describe its physics. In Section~\ref{methods} we provide details of the various methodologies used in our study---both experimental and computational. We then report our results, beginning with the effect of pressure at ambient temperature, and proceeding thereafter to consider the way in which this behaviour varies as temperature is reduced. The paper concludes with a discussion of the implications of our findings and a summary of the most important outstanding questions that remain to be addressed in future studies.

\section{cadmium cyanide: background}

\begin{figure}
\includegraphics{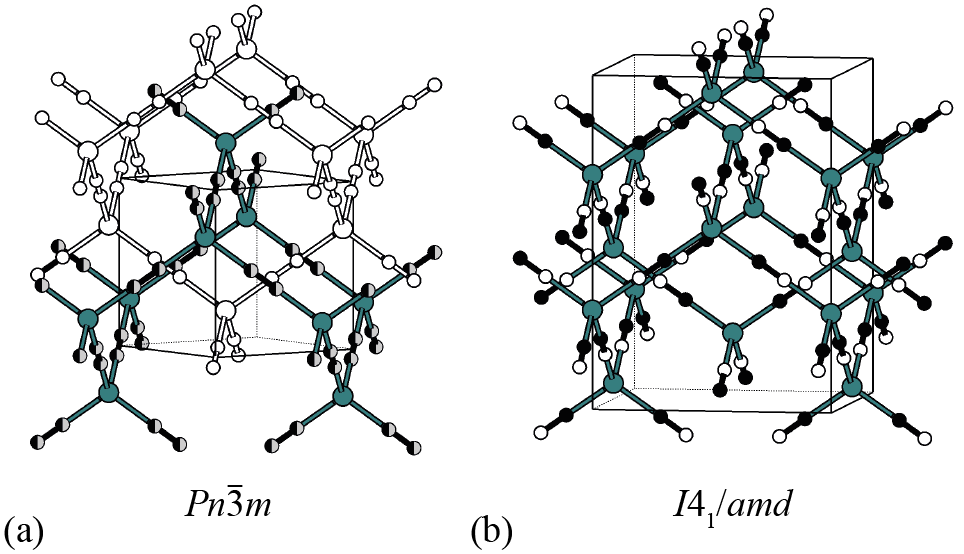}
\caption{\label{fig1} (a) Representation of the cubic structure of ambient-phase Cd(CN)$_2$-I. Cd atoms are shown in teal, and C/N atoms (which are disordered) in dual-tone black/gray. One of the two interpenetrating diamondoid nets is coloured white. (b) The tetragonal Cd(CN)$_2$ structure that emerges at ambient pressure for temperatures below about 135\,K. C and N atoms are now shown in black and white, respectively; note the presence of cyanide-ion orientational order. The two frameworks have slipped relative to one another along the tetragonal (vertical) axis.}
\end{figure}

At room temperature and ambient pressure, Cd(CN)$_2$ adopts the anticuprite structure type with $Pn\bar3m$ crystal symmetry [Fig.~\ref{fig1}(a)] \cite{Shugam_1945,Goodwin_2005,Fairbank_2012,Coates_2021}. Cd centres are tetrahedrally coordinated by cyanide ions, and each cyanide ion bridges a pair of Cd$^{2+}$ ions. These Cd--C--N--Cd linkages connect to form a network with the diamond topology; two such nets interpenetrate. The compounds Zn(CN)$_2$ and Mn(CN)$_2$ are isostructural to Cd(CN)$_2$ \cite{Zhdanov_1941,Williams_1997,Manson_1998}. All three systems show head-to-tail disorder of the cyanide ion orientations \cite{Nishikiori_1991,Hibble_2013,Coates_2021}.

As is the case for Zn(CN)$_2$ (and, presumably, Mn(CN)$_2$), the cubic unit cell of Cd(CN)$_2$ expands on cooling \cite{Goodwin_2005}. This NTE effect is quantified by the volumetric coefficient of thermal expansion, defined as the relative rate of change in volume at constant pressure:
\begin{equation}
\alpha_V=\frac{1}{V}\left(\frac{\partial V}{\partial T}\right)_p.
\end{equation}
Variable-temperature powder X-ray diffraction measurements give the value $\alpha_V=-55(2)$\,MK$^{-1}$ for Cd(CN)$_2$ at ambient pressure over the temperature range 150--750\,K \cite{Coates_2021b}. For context, the corresponding values for the high-profile isotropic NTE materials ScF$_3${\color{black}, CaZrF$_6$,} and ZrW$_2$O$_8$ are $-$12\,MK$^{-1}$ (10--1100\,K){\color{black}, $-$21.5(7)\,MK$^{-1}$,} and $-26.1$\,MK$^{-1}$ (0.3--1050\,K), respectively \cite{Mary_1996,Greve_2010,Hancock_2015}.

Gr{\"u}neisen theory provides a semi-quantitative rationalisation for NTE in Cd(CN)$_2$ in terms of thermal population of volume-reducing phonon modes. For each branch of the phonon spectrum $\nu$ and for each wave-vector $\mathbf q$, the mode Gr{\"u}neisen parameter
\begin{equation}
\gamma(\nu,\mathbf q)=-\frac{{\rm d}[\ln\omega(\nu,\mathbf q)]}{{\rm d}(\ln V)}\label{gamma}
\end{equation}
is a dimensionless quantity that captures the coupling between phonon frequency $\omega$ and crystal volume $V$ \cite{Gruneisen_1926,Barrera_2005}. Intuitively, one expects vibrational frequencies to increase with decreasing volume (shorter bonds = stiffer bonds), and hence $\gamma(\nu,\mathbf q)>0$. NTE modes, by contrast, are vibrations for which $\gamma<0$: like a plucked guitar-string, their frequency drops as the system is compressed. The bulk {\color{black} Gr{\"u}neisien} parameter $\gamma(T)$ integrates over all modes and all periodicities, with the contribution of each $\gamma(\nu,\mathbf q)$ weighted by the corresponding contribution of the mode to the specific heat at temperature $T$. Hence $\gamma(T)$ is negative if and only if there is a large number of individual $\gamma(\nu,\mathbf q)<0$ terms for modes with low frequencies. The explicit link to NTE comes {\it via} the relationship
\begin{equation}
\alpha_V=\frac{\gamma C_V}{VB},
\end{equation}
where $B$ is the isothermal bulk modulus and $C_V$ the specific heat at constant volume. Since $V$, $B$, and $C_V$ are all strictly positive, phonon-mediated isotropic NTE implies a negative value of $\gamma$. For Cd(CN)$_2$, the Gr{\"u}neisen parameters calculated using \emph{ab initio} density functional theory (DFT) methods adopt a range of values between $-$20 and 2 and collectively give $\alpha_V=-39$\,MK$^{-1}$ \cite{Zwanziger_2007,Ding_2008}.

Ultimately the vibrational modes responsible for NTE---\emph{i.e.} those with large and negative Gr{\"u}neisen parameters---appear to involve transverse displacements of CN$^-$ ions away from the Cd$\ldots$Cd axes, as envisaged in Refs.~\citenum{Goodwin_2005,Goodwin_2006} and implicated in other `tension-effect' NTE materials \cite{Evans_1999}.

An important early indication that the physics of Cd(CN)$_2$ might be more complicated than implied by this relatively conventional picture was the experimental observation of a displacive phase transition on cooling below {\color{black}135}\,K that occurs despite the absence of any phonon instabilities identified in DFT calculations \cite{Goodwin_2005,Fairbank_2012,Zwanziger_2007}. We have recently shown that this transition is driven by temperature-dependent ordering of CN$^-$ ion orientations, and hence there is strong coupling between long-timescale CN$^-$ ion `flips' and the elastic behaviour of Cd(CN)$_2$ \cite{Coates_2021}. The thermal accessibility of CN$^-$ flips distinguishes the behaviour of Cd(CN)$_2$ from that of its better-studied analogue Zn(CN)$_2$. Cyanide orientations are fixed in the latter \cite{WernerZwanziger_2012,Hibble_2013} and its ambient-temperature $Pn\bar3m$ structure is stable down to the very lowest temperatures for which crystallographic data have been measured (25\,K) \cite{Goodwin_2005}.

The low-temperature structure of Cd(CN)$_2$, as determined using neutron scattering measurements, has tetragonal $I4_1/amd$ symmetry [Fig.~\ref{fig1}(b)] \cite{Coates_2021}. This structure differs from that of the ambient $Pn\bar3m$ phase in two key respects. First, the cyanide orientations are no longer disordered, but are progressively arranged such that each Cd centre is coordinated by two C and two N atoms. This allows the Cd atom to displace away from the centre of the CdC$_2$N$_2$ coordination tetrahedron, giving rise to a local dipole \cite{Fairbank_2012,Coates_2021}. Within each of the two interpenetrating diamondoid nets these {\color{black} Cd-displacement} dipoles point along a single common direction (parallel to one of the $\langle100\rangle$ axes of the original cubic cell), but the dipole orientation inverts from one net to the other to give antiferroelectric cyanide order. The second key difference is that the two interpenetrating diamondoid networks---evenly spaced in the ambient $Pn\bar3m$ structure---are displaced in opposite directions along the tetragonal axis at low temperatures. The relationship between ambient- and low-temperature Cd(CN)$_2$ structures is similar to that between the high-pressure VII and VIII phases of water ice \cite{Kuhs_1984,Jorgensen_1984,Kuo_2004}.

In Ref.~\citenum{Coates_2021} we rationalised this particular CN$^-$ ordering transition in terms of a simple pseudospin model of CN$^-$ dipolar degrees of freedom. Denoting the orientation of each CN$^-$ ion by a classical vector $\mathbf S$, we found that the `spin-ice'-like Hamiltonian
\begin{equation}\label{hamil}
\mathcal H = -J_{\rm eff}\sum_{i,j}\mathbf S_i\cdot\mathbf S_j+D\sum_{i,j^\prime}\mathbf S_i\cdot\mathbf S_{j^\prime}-\Delta\sum_iS_{i\parallel}^2
\end{equation}
actually captured the key physics of CN$^-$ order and disorder in Cd(CN)$_2$ surprisingly well. Here, $J_{\rm eff}$, $D$, and $\Delta$ correspond respectively to an effective interaction between nearest-neighbour CN$^-$ ions in the same network, an interaction between nearest-neighbour CN$^-$ ions in alternate networks, and a `single-ion' term that reflects the barrier height to CN$^-$ flips. Experiment and theory converge on essentially the same set of values $J_{\rm eff}=205$\,K, $D=85$\,K, $\Delta=8\,800$\,K, for which Eq.~\eqref{hamil} predicts a phase transition at $T_{\rm c}=121$\,K to precisely the $I4_1/amd$ structure type observed experimentally. Since the CN$^-$ flipping mechanism involves a decrease in Cd$\ldots$Cd separation, one expects some interplay between CN$^-$ flips and NTE, and again between these flips and the behaviour of Cd(CN)$_2$ under pressure.

Whereas the effect of pressure on the structural and spectroscopic behaviour of Zn(CN)$_2$ has been relatively extensively studied \cite{Dehnicke_1974,Chapman_2007,Ravindran_2007b,Mittal_2009,Mittal_2011,Collings_2013,Lapidus_2013,Fang_2013b}, to the best of our knowledge the only variable-pressure investigation of Cd(CN)$_2$ is the early infrared study of Ref.~\citenum{Dehnicke_1974}. That work identified a single phase transition at pressure $<3$\,GPa in which the Cd environment appeared to increase in symmetry. Any attempts to determine the structural evolution of Cd(CN)$_2$ using variable-pressure X-ray diffraction measurements will have suffered from the extreme sensitivity to X-ray damage that severely complicate reliable structure determination \cite{Coates_2021b}.

\section{Methods}\label{methods}

Here we investigate the high-pressure behaviour of Cd(CN)$_2$ using a variety of experimental and computational approaches. In this section we provide details of the specific methods used in our study.

\subsection{Synthesis}
A sample of isotopically enriched Cd(CN)$_2$ was prepared according to the method outlined in Ref.~\citenum{Coates_2018}. Hg(CN)$_2$ (1\,g, Aldrich, 99\%) and excess $^{114}$Cd metal (1\,g, Isoflex, 99.1(3)\% $^{114}$Cd) were added to one arm of a custom-made glass N-cell. Anhydrous ammonia gas (30\,mL liquid ammonia volume, BOC) was condensed onto the mixture and stirred for six hours in an acetone/dry-ice bath with the temperature maintained between 240 and 250\,K. The mixture was filtered---through the porous frit separating the two Schlenk tubes of the N-cell---under flowing ammonia gas to remove insoluble Hg. The resulting solution allowed to evaporate, yielding a polycrystalline sample of $^{114}$Cd(NH$_3)_2[^{114}$Cd(CN)$_4$]. This was then heated at 80\,$^{\circ}$C for 24\,hours to yield $^{114}$Cd(CN)$_2$. The sample used here is the same as that on which the earlier neutron scattering study of Ref.~\citenum{Coates_2021} was based.

\subsection{Variable-temperature/variable-pressure neutron diffraction measurements}
Variable-temperature/variable-pressure neutron diffraction measurements were carried out using the HRPD and OSIRIS instruments at the ISIS spallation source.

For the HRPD measurements, 1.3658\,g of $^{114}$Cd(CN)$_2$ was loaded into a TiZr gas pressure cell, which was then pressurised with argon gas as the pressure-transmitting medium. The hydrostatic pressure was generated by an external pressure intensifier unit to an accuracy of 1 bar. Ar was chosen instead of He {\color{black} through an abundance of caution} to avoid penetration of the smaller He atoms inside the Cd(CN)$_2$ framework structure. The temperature was controlled using a closed cycle refrigerator (CCR) between 300 and 100\,K. Data were collected on increasing pressure between 0 and 0.5\,GPa in 0.1\,GPa increments. The initial pressure was 50\,bar (0.005\,GPa) to ensure that the pressure cell was sealed. At each pressure point data were collected on cooling at five temperatures (300, 250, 200, 150 and 100\,K), after which the temperature was raised to 300\,K and the pressure increased. Data collections of 80\,$\mu$A ($\sim$2\,hours) were carried out at each point, using the instrument's standard 30--130\,ms time-of-flight window. Due to the substantial incoherent background produced by the TiZr alloy, which overwhelms the coherent scattering from the sample in backscattering geometry, only data from HRPD's 90-degree detector banks were used for profile refinement. The time-of-flight bands stated above correspond to $d$-spacing ranges of 0.85--3.90\,\AA\ in the 90-degree detectors.

For the OSIRIS measurements, 1.300\,g of $^{114}$Cd(CN)$_2$ was loaded into a TAV-6 (Ti-6Al-4V alloy) gas pressure cell. Unlike the TiZr cell, TAV-6 is not null scattering, but contains well-defined Bragg peaks at low $d$-spacing and has a negligible incoherent background signal \cite{Halevy_2010}. An additional advantage of this cell is that the greater strength of the TAV-6 alloy relative to TiZr allows the pressure vessel's walls to be thinned from 10.5 to 5.3\,mm, whilst still achieving a maximum working pressure of 0.54\,GPa. This greatly reduces the attenuation of the incident beam and of the diffracted rays. The cell was then pressurised using Ar gas as the pressure-transmitting medium up to a minimum starting pressure of 10\,bar (0.001\,GPa). Data were collected using two separate time-of-flight bands in order to access an appropriate $d$-spacing range in the instrument's only {\color{black} available} diffraction detectors, which are positioned in backscattering geometry. The corresponding time-of-flight bandwidths were 29.4--69.4\,ms and 47.1--87.1\,ms, giving $d$-ranges of 1.8--4.0\,\AA\ and 2.9--4.9\,\AA, respectively.  The experimental protocol was chosen in order to maximise the number of data points within the ambient $Pn\bar3m$ phase. Data were collected in intervals of 0.02\,GPa from 0--0.2\,GPa at 300\,K, before cooling to 275\,K (at pressure). The gas cell was then depressurised after a period of two hours (the time deemed necessary for the induced activity of Ar to decay to a level at which the gas could be safely vented). Once at 275\,K, data were again collected in intervals of 0.02\,GPa up to 0.2\,GPa before cooling to 250\,K at pressure, and then depressurising to 0.005\,GPa. Again, data were measured in 0.02\,GPa increments up to 0.18\,GPa at 250\,K, at which point there appeared to be evidence of sample history dependence. The sample was then warmed to 300\,K at 0.2\,GPa before depressurising at 300\,K, then cooling to 250\,K, at which point the pressure points 0-0.2\,GPa were repeated in 0.02\,GPa increments. 

Pawley refinements were carried out using TOPAS Academic v.4.1 \cite{TOPAS} to obtain the $T$/$p$ dependence of the cubic lattice parameter of Cd(CN)$_2$. The instrument-dependent peak shape parameters were refined against diffraction data for a Si 640c NIST standard. The Bragg peaks corresponding to the TAV-6 pressure cell were refined as two Pawley phases, the first corresponding to the structure of hcp-Ti with space group symmetry $P6_3/mmc$ ($a\sim2.924$\,\AA, $c\sim4.670$\,\AA) and a second fcc phase with $Fm\bar3m$ symmetry ($a\sim4.532$\,\AA). Lattice parameters used in the equation of state for Cd(CN)$_2$ were obtained using Pawley refinements. For the OSIRIS data, the $d$-range used for Pawley refinements was confined to $3<d<4.6$\,\AA. Three peaks corresponding to the (002), (111) and (011) $Pn\bar3m$ Bragg reflections were measured in this region at all pressures. The signal at lower $d$-spacings was dominated by strong reflections from the TAV-6 sample container, and so this region was not used for Pawley refinements for lattice parameter determination. 

\subsection{Variable-pressure inelastic neutron scattering measurements}

Variable-pressure inelastic neutron scattering experiments were performed at ambient temperature on the same $\sim1$\,g of $^{114}$Cd(CN)$_2$ described above using the MARI instrument at ISIS. The incident neutron energy was 10\,meV (400\,Hz choppers), which gave an energy resolution of about 0.175\,meV at an energy transfer of 2\,meV.  Pressure was applied using {\color{black} the same TAV-6 cell used on the OSIRIS beamline, chosen because of its low absorption and background scattering that can often complicate high-pressure inelastic neutron scattering experiments. For these measurements, we made use of He as a pressure-transmitting medium rather than Ar, in order to reduce complications associated with activation of cycled gas. We will come to show there is in fact no evidence of He intercalation within the structure.} Each measurement corresponded to a neutron beam exposure of approximately 1000\,$\mu$A\,h and was corrected for background scattering from the sample environment. 

\subsection{Quantum mechanical calculations}
\emph{Ab initio} lattice dynamical calculations were performed using the PAW formalism \cite{Blochl_1994} as implemented in VASP \cite{Kresse_1993, Kresse_1996, Kresse_1996b}. We used PBE functionals with a plane-wave cut-off of 1600\,eV to avoid Pulay stress \cite{Perdew_1996,Pulay_1969}. Electronic self-consistent field cycles were converged to 10$^{-6}$\,eV and the convergence criteria for the total energy and ionic forces were set to 10$^{-8}$\,eV and 10$^{-6}$\,eV\,\AA$^{-1}$, respectively. The structures were allowed to fully relax---ions and lattice parameters were optimised simultaneously---for different representative Cd(CN)$_2$ models to account for varying degrees of cyanide order. These relaxed cells were then used as the basis for elastic property calculations. In each case, we generated a series of seven structures with expanded/contracted cells $-0.5<\Delta V/V<+0.5\%$. The volume dependence of the energy was fitted using a third-order Birch-Murnaghan equation of state in order to determine the zero-pressure bulk modulus $B_0$ and its first pressure derivative $B_0^\prime$ \cite{Hebbache_2004}. Phonon dispersion relations were calculated using the \textsc{Phonopy} package \cite{Togo_2015}, and thermal expansivities determined within the quasi-harmonic approximation using \textsc{Phonopy}-qha \cite{Togo_2010}.

\section{Results and Discussion}

\subsection{Cd(CN)$_2$-I equation of state}

\begin{figure}
\includegraphics{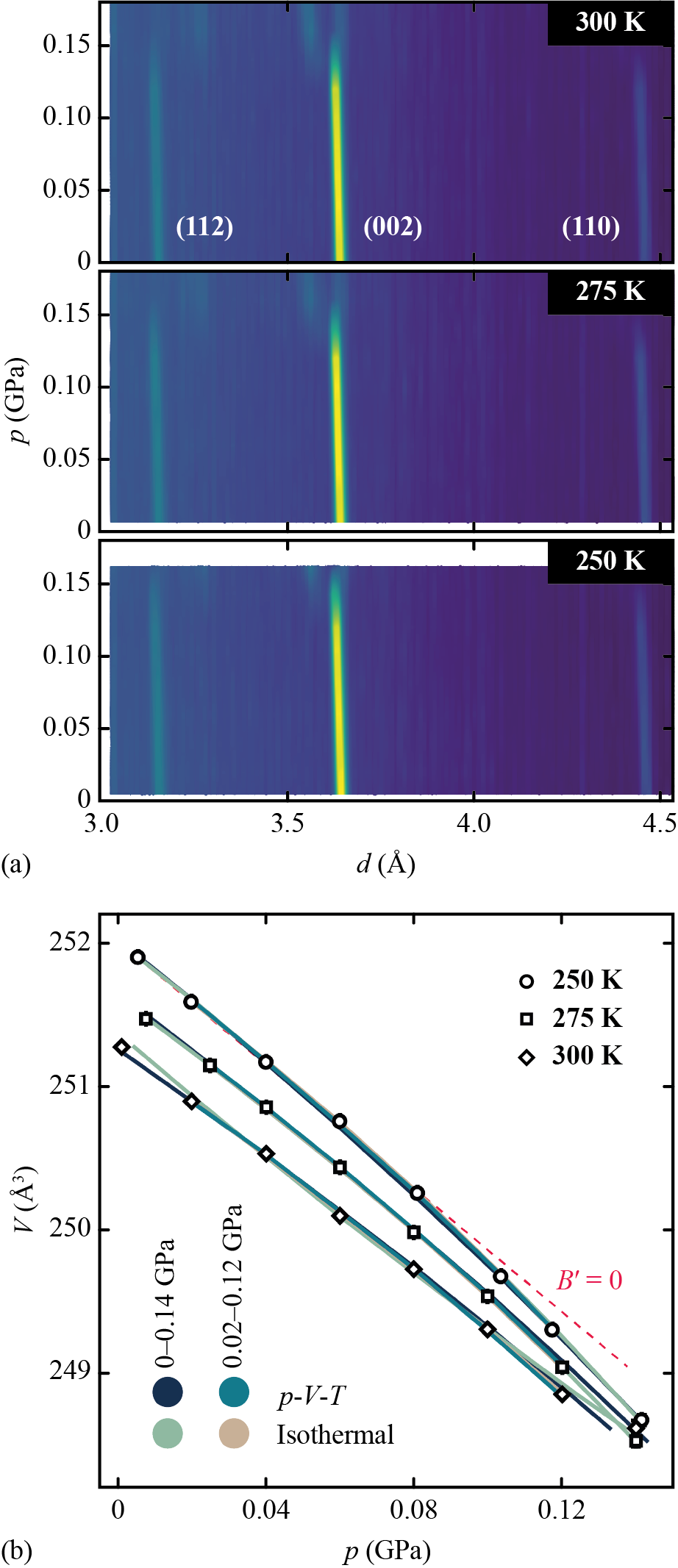}
\caption{\label{fig2}{\color{black}(a) Variable-pressure neutron diffraction patterns of Cd(CN)$_2$-I at 300, 275 and 250\,K. At each temperature there is a first-order transition to a lower-symmetry phase at a critical pressure of about 0.14\,GPa. (b) Equation of state of Cd(CN)$_2$-I, with data extracted {\it via} Pawley refinement against the diffraction patterns shown in (a). The solid curves correspond to fits obtained using Eq.~\eqref{bmt} ($p$--$V$--$T$) or a third-order Birch--Murnaghan equation of state as discussed in the text. The dashed line shows a linear limit ($B_0^\prime=0$) for the 250\,K data, so as to emphasise the extreme pressure-induced softening observed at this temperature.}}
\end{figure}

Our starting point was to explore the structural behaviour of Cd(CN)$_2$ under compression at ambient temperature. Fig.~\ref{fig2}(a) shows the evolution of a key region of the neutron diffraction pattern as a function of pressure, as measured using the OSIRIS instrument. From these data, it is clear that the ambient-pressure phase is stable only up to a pressure of about 0.14\,GPa whereupon it undergoes a first-order transition to a high-pressure phase. We hereafter use the labels Cd(CN)$_2$-I and Cd(CN)$_2$-II to denote the ambient and high-pressure phases. Pawley refinements against the neutron diffraction patterns measured at varying pressure values allowed extraction of the $V(p)$ equation of state, which is shown in Fig.~\ref{fig2}(b). We used the third-order Birch Murnaghan formalism \cite{Birch_1947,Sata_2002}
\begin{eqnarray}\label{3bm}
p(V)& = & \frac{3B_{0}}{2}\left[\left(\frac{V_{0}}{V}\right)^{\frac{7}{3}}-\left(\frac{V_{0}}{V}\right)^{\frac{5}{3}}\right]\nonumber\\
&&\times \left\{1+\frac{3}{4}(B_0^{\prime}-4)\left[\left(\frac{V_{0}}{V}\right)^{\frac{2}{3}}-1\right]\right\}
\end{eqnarray}
to fit these data and so estimate the zero-pressure bulk modulus $B_0$ and its pressure dependence $B_0^\prime$. In this way, we obtained $B_0=13.6(4)$\,GPa and $B_0^\prime=-18(5)$. The corresponding value of $V_0$ was 251.271(19)\,\AA$^3$.

We put these values in context by contrasting the pressure-dependent behaviour of Cd(CN)$_2$ with that of its better-studied close relative Zn(CN)$_2$. There are a few key differences. First, the former is about three times softer than the latter ($B_0 =36.9(22)$\,GPa in Zn(CN)$_2$) \cite{Chapman_2007,Collings_2013}. While both materials become softer as pressure is increased---the anomalous `pressure-induced softening' effect characteristic of NTE materials \cite{Fang_2013c,Fang_2013b}---Cd(CN)$_2$ does so about three times as quickly as Zn(CN)$_2$ (for which $B_0^\prime=-6.0(7)$) \cite{Chapman_2007,Collings_2013}. So Cd(CN)$_2$ is elastically very soft, and becomes even softer very quickly under pressure. This description is consistent with our observation of a pressure-induced phase transition at remarkably low pressures: indeed Cd(CN)$_2$-I is about an order of magnitude less stable to compression than Zn(CN)$_2$-I, which persists to $\sim$1.5\,GPa \cite{Collings_2013}.

Since the Cd(CN)$_2$-I structure is known to be stable at ambient pressure on cooling to $\sim${\color{black}135}\,K, we repeated our isothermal high-pressure neutron diffraction measurements at reduced temperatures of 275 and 250\,K. The corresponding data are shown in Fig.~\ref{fig2}(a). We observed qualitatively similar behaviour at both temperatures to that observed at 300\,K---in that the ambient phase is stable only up to pressures of about 0.14\,GPa. The critical pressure appeared to show very little temperature dependence, and we noted from diffraction measurements performed on decompression (data not shown) a hysteresis in the Cd(CN)$_2$-I/II transition. The corresponding unit-cell volumes were again extracted from our variable-pressure diffraction patterns using Pawley refinements, and we show the corresponding data in Fig.~\ref{fig2}(b). As expected for an NTE material, the unit-cell volume increases on cooling. Also evident is an increased pressure-induced softening effect as temperature is reduced.

\begin{table*}
\center
  \caption{{\color{black}Thermodynamic parameters derived from the different equation of state fits for Cd(CN)$_2$ shown in Fig.~\ref{fig2}(b). The two equations of state involved are the isothermal third-order Birch--Murnaghan (Eq.~\eqref{3bm}, `Isothermal') and that given by Eq.~\eqref{bmt} (`$p$--$V$--$T$').}}
  \label{tabley}
{\color{black}
 \begin{tabular}{lccccccc}
   \hline
\multirow{2}{*}{EoS}&$p$ &$T$&$B_0$&$B_0^\prime$&$V_0$&$\alpha_V$ &$\displaystyle\left(\frac{\partial B}{\partial T}\right)_p$ \\
&(GPa)&(K)&(GPa)&&\AA$^3$&MK$^{-1}$&(GPa\,K$^{-1}$)\\ \hline
\multirow{2}{*}{$p$--$V$--$T$}&0--0.14&250--300&13.0(4)&$-$18(4)&251.64(2)&$-$61(3)&0.038(6)\\
&0.02--0.12&250--300&13.5(2)&$-$26(3)&251.624(14)&$-$60.0(13)&0.030(3)\\ \hline
\multirow{7}{*}{Isothermal}&\multirow{3}{*}{0--0.14}&300&12.2(6)&6.1(8)&251.31(3)&--&--\\
&&275&13.5(5)&$-$26(5)&251.60(3)&--&--\\
&&250&12.6(4)&$-$23(5)&252.01(3)&--&--\\ 
&\multirow{3}{*}{0.02--0.12}&300&13.6(4)&$-$18(5)&251.271(19)&--&--\\
&&275&13.4(4)&$-$25(4)&251.62(3)&--&--\\
&&250&13.4(3)&$-$32(3)&251.97(2)&--&--\\ 
    \hline \hline
  \end{tabular}
}\end{table*}

To quantify the combination of both thermal and pressure effects, we {\color{black}have used two approaches. On the one hand, we fitted isothermal data sets using the conventional third-order Birch-Murnaghan expression of Eq.~\eqref{3bm}, obtaining distinct values of $V_0$, $B_0$ and $B_0^\prime$ at each temperature. And, on the other hand, we also used a modified version of the same expression
\begin{eqnarray}\label{bmt}
p(V,T)& = & \frac{3B_{0}(T)}{2}\left[\left(\frac{V_{0}(T)}{V}\right)^{\frac{7}{3}}-\left(\frac{V_{0}(T)}{V}\right)^{\frac{5}{3}}\right]\nonumber\\
&&\times \left\{1+\frac{3}{4}(B_0^{\prime}-4)\left[\left(\frac{V_{0}(T)}{V}\right)^{\frac{2}{3}}-1\right]\right\},\label{pvt}
\end{eqnarray}
where
\begin{equation}
V_{0}(T)=V_0(T_{\rm ref})[1+\alpha_V(T-T_{\rm ref})]
\end{equation}
and
\begin{equation}
B_0(T)=B_0(T_{\rm ref}) +\left(\frac{\partial B}{\partial T}\right)_p(T-T_{\rm ref}).
\end{equation}
The purpose of this second approach is to include the lowest-order terms that allow for variation in system volume and stiffness as a function of temperature. Given the additional uncertainty of pressure determination at the lowest and highest pressure values, we repeated our analysis over both the full (0--0.14\,GPa) and reduced (0.02--0.12\,GPa) pressure intervals. We used the intermediate temperature value (275\,K) as $T_{\rm ref}$, and fitted Eq.~\eqref{pvt} to the ensemble of variable-$p/T$ unit-cell data measured at 250, 275, and 300\,K. The corresponding fits are shown in Fig.~\ref{fig2}(b) and the associated thermodynamic parameters are listed in Table~\ref{tabley}. Despite the variety of fits used, the key parameters show remarkable consistency.

The parameter $\left(\frac{\partial B}{\partial T}\right)_p$ quantifies the thermal variation in material stiffness and deserves particular comment.} That the value obtained is positive for Cd(CN)$_2$ implies that the bulk modulus increases with temperature. This effect---termed \emph{{\color{black}thermal} hardening}---is counterintuitive because materials normally soften on heating \cite{Dove_2016}. Its origin in NTE materials is thought to involve an increased anharmonicity of NTE modes at higher temperatures {\color{black}\cite{Fang_2013,Morelock_2013,Dove_2016}}. Thermodynamically, the quantity is related to the isothermal pressure dependence of the expansivity: \cite{Dove_2016}
\begin{equation}\label{warmhard}
\left(\frac{\partial\alpha_V}{\partial p}\right)_T = \frac{1}{B^2} \left(\frac{\partial B}{\partial T}\right)_p.
\end{equation}
So, we find that the magnitude of NTE in Cd(CN)$_2$ is actually reduced under pressure---\emph{i.e.}\ $(\frac{\partial\alpha_V}{\partial p})_T>0$. This observation again runs counter to intuition: one might reasonably expect that, since NTE modes decrease in energy with applied pressure, NTE should be \emph{enhanced} as pressure is increased. Zn(CN)$_2$ is conventional in this respect, as its room-temperature NTE effect increases in magnitude from ambient pressure ($\alpha_V=-52.2(5)$\,MK$^{-1}$) to 0.4\,GPa ($\alpha_V=-58.2(6)$\,MK$^{-1}$) \cite{Chapman_2007}. The anomalous behaviour of Cd(CN)$_2$ again reflects the argument made in Refs.~\citenum{Fang_2013,Dove_2016} that anharmonic effects---if sufficiently strong---can invert this relationship.

\subsection{Inelastic neutron scattering}

\begin{figure}
\includegraphics{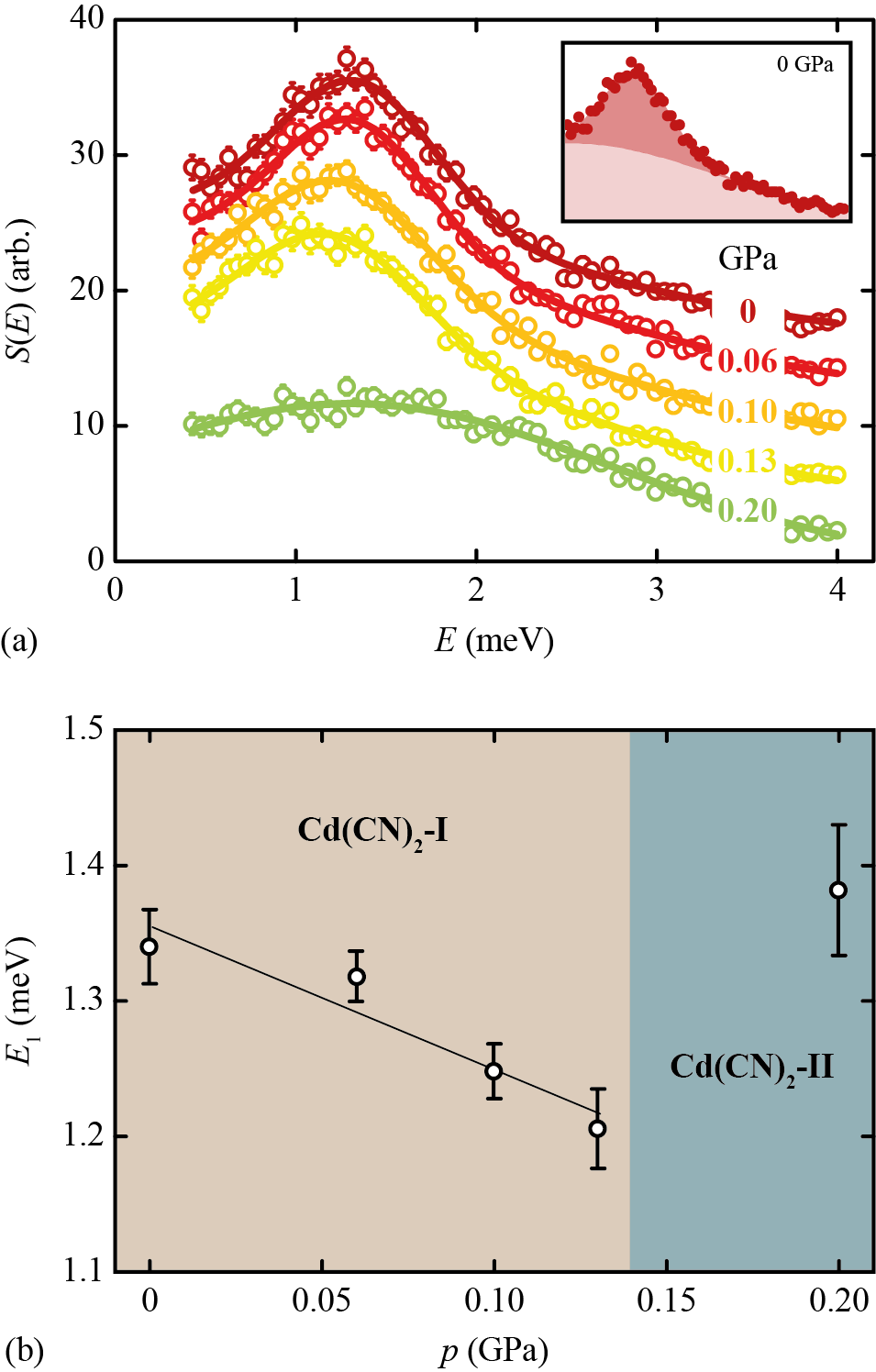}
\caption{\label{fig3}(a) Neutron-weighted low-energy phonon densities of states in Cd(CN)$_2$ measured using the MARI spectrometer {\color{black} at ambient temperature}. Data are shown as open circles, and fits using Eq.~\eqref{gaussfit} are shown as solid lines. Successive curves have been shifted vertically by four units to aid interpretation. {\color{black}The inset shows the contributions of the two Gaussian components (darker = narrower, index 1; lighter = broader, index 2) to the fit for the 0 GPa data.} (b) Pressure dependence of the effective mean value {\color{black}$E_1$} of the low-energy phonon dispersion, obtained from the fits to data shown in (a). The decrease in {\color{black}$E_1$} with pressure is characteristic of NTE materials, for which {\color{black} $\gamma_1$} is negative. {\color{black}The value of $\gamma_1$ was determined by an error-weighted least-squares linear fit to these data, shown here as a thin line.} The NTE modes appear to stiffen on progression into the high-pressure phase Cd(CN)$_2$-II.}
\end{figure}

In order to provide further insight into anharmonicity in Cd(CN)$_2$-I, we used inelastic neutron scattering measurements to monitor the effect of pressure on the neutron-weighted phonon density of states. We carried out measurements at five pressure values: 0, 0.06, 0.1, 0.13, and 0.2\,GPa. The $Q$-integrated $S(E)=\int S(Q,E)\,{\rm d}Q$ inelastic neutron scattering function is shown in Fig.~\ref{fig3}(a) for the low-energy domain $0\leq E\leq 4$\,meV. This energy range includes the relatively dispersionless phonon branches understood to dominate NTE response \cite{Zwanziger_2007}. Indeed, at all pressures, $S(E)$ is dominated by a single feature centred at $E\simeq1.2$\,meV. For pressures below $p_{\rm c}\simeq0.14$\,GPa, the $S(E)$ function is very similar, with the position of the main feature shifting to lower energy as pressure increases. By contrast, the $S(E)$ function measured for Cd(CN)$_2$-II at 0.2\,GPa is very different: while there is still a single peak, it is much broader and centred at a higher $E$. {\color{black}This change in behaviour between 0.13 and 0.2\,GPa is of course consistent with the phase transition observed in our diffraction experiments. That the critical pressure involved is the same between the two experiment types (the diffraction measurements performed with Ar as pressure transmitting medium and the inelastic scattering with He) indicates that there is no evidence for He intercalation in Cd(CN)$_2$---at least over the pressure ranges explored here.}

We have quantified these various observations by fitting the $S(E)$ functions using a double Gaussian {\color{black}
\begin{eqnarray}\label{gaussfit}
S(E)&\simeq&\frac{c_1}{\sqrt{2\pi}w_1}\exp\left[-\frac{(E-E_1)^2}{w_1^2}\right]\nonumber\\ &&+\frac{c_2}{\sqrt{2\pi}w_2}\exp\left[-\frac{(E-E_2)^2}{w_2^2}\right],
\end{eqnarray}
where the amplitudes $c_i$, widths $w_i$ and position $E_i$ were treated as free parameters. We constrained the width $w_2$ and midpoint $E_2$ of the broader Gaussian component to be constant for all datasets. We show in Fig.~\ref{fig3}(b) the pressure-dependence of the peak position $E_1$, which reflects the various observations noted above; the corresponding parameters, obtained through a single 22-parameter least-squares fit to the full experimental data set, are given in Table~\ref{tablex}. Interpreting the value of $E_1$} as an approximation to the mean energy of the low-frequency phonon branches in Cd(CN)$_2$-I, we use the relation
{\color{black}
\begin{equation}
\gamma_1=\frac{B_0}{E_1}\frac{{\rm d}E_1}{{\rm d}p},
\end{equation}
derived from Eq.~\eqref{gamma}, to obtain the estimate $\gamma_1\simeq-11(3)$. Such a large and negative Gr{\"u}neisen parameter is consistent with the value $\gamma=-20$ determined for the lowest-energy NTE modes in Cd(CN)$_2$ using density functional theory calculations \cite{Zwanziger_2007}, and comparable to those measured for other NTE materials, such as ZrW$_2$O$_8$ \cite{Mittal_2001}.}

\begin{table}
\center
  \caption{{\color{black}Pressure-dependence of the variable parameters used in Eq.~\eqref{gaussfit} to obtain the fits shown in Fig.~\ref{fig3}. In all cases, $E_2=0.525$\,GPa and $w_2=2.535$\,GPa. All values have units of GPa.}}
  \label{tablex}
{\color{black}
 \begin{tabular}{lcccc}
   \hline
Pressure&$E_1$&$c_1$&$w_1$&$c_2$\\ \hline
0&1.340&15.385&0.607&65.864\\
0.06&1.318&14.058&0.587&77.438\\
0.10&1.248&14.812&0.678&78.049\\
0.13&1.206&14.016&0.685&81.519\\
0.20&1.382&49.122&1.970&11.971\\
    \hline \hline
  \end{tabular}
}\end{table}

\subsection{Exploration of the wider $p/T$ phase diagram}

\begin{figure}[b]
\includegraphics{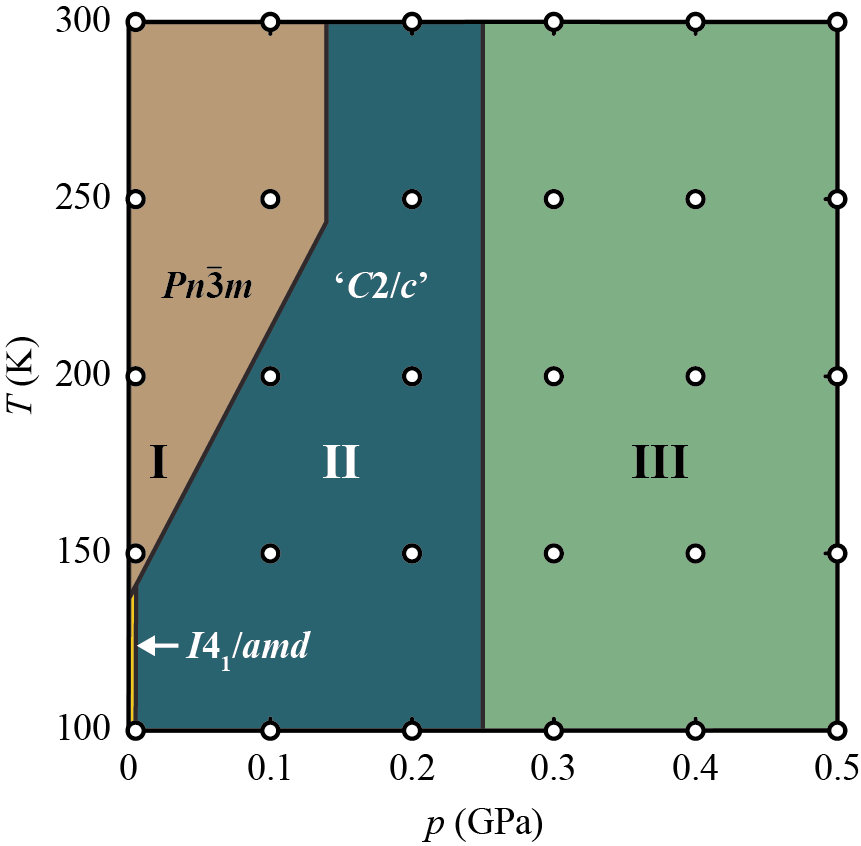}
\caption{\label{fig4}Pressure/temperature phase diagram of Cd(CN)$_2$ as determined using neutron diffraction measurements. The grid of $(p,T)$ values interrogated using HRPD are shown as open white circles. This diagram includes the low-temperature ambient-pressure $I4_1/amd$ phase reported in Ref.~\citenum{Coates_2021}.}
\end{figure}

Variable-temperature / variable-pressure neutron diffraction measurements (HRPD instrument) were used to map the phase behaviour of Cd(CN)$_2$ over a larger temperature and pressure domain. A total of 30 different $p,T$ values were interrogated, spanning the ranges $0.005\leq p\leq0.5$\,GPa and $100\leq T\leq300$\,K [Fig.~\ref{fig4}]. These $p,T$ values were evenly spaced, and we note that the minimum pressure 0.005\,GPa corresponds to that required to seal the gas cell used in our measurements. In total we identified three distinct phases distributed across the $p,T$ values we sampled, giving rise to the phase diagram shown in Fig.~\ref{fig4}.

\begin{figure}
\includegraphics{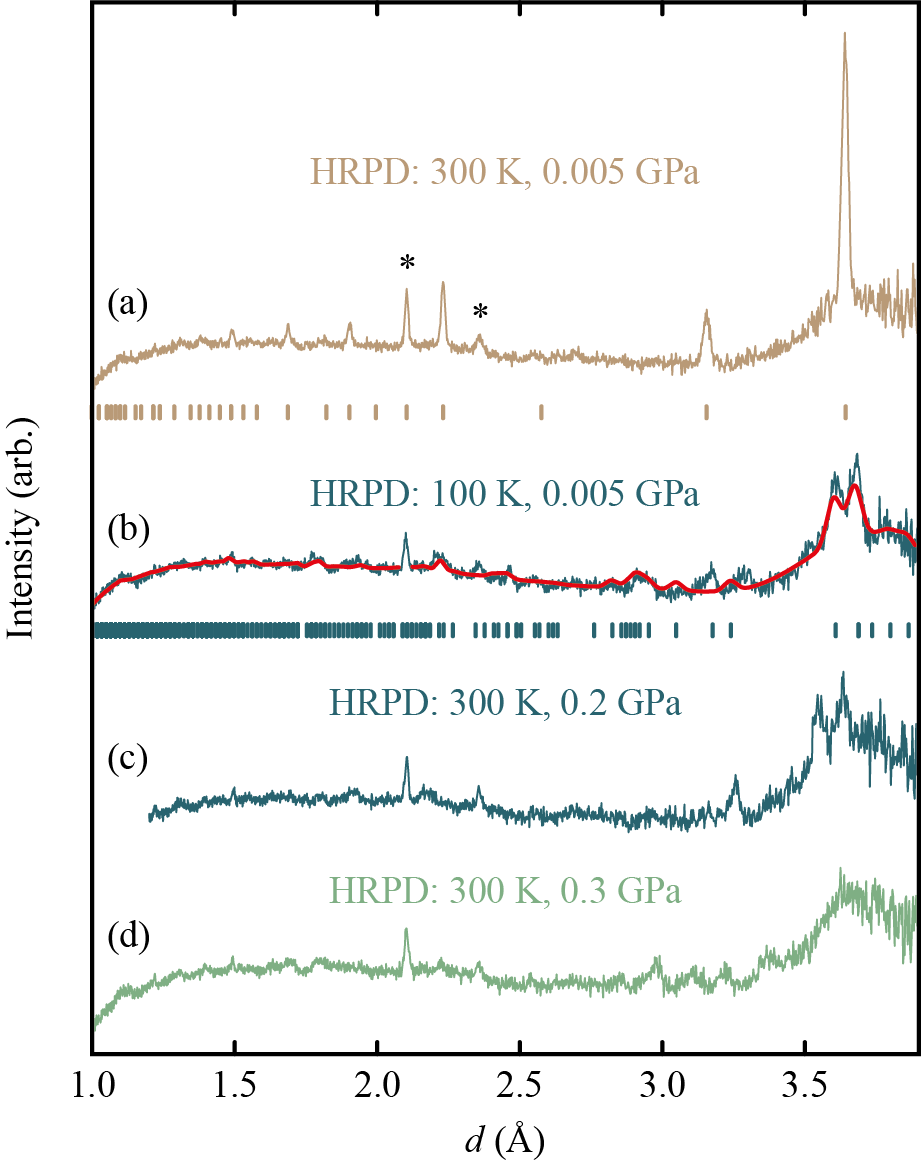}
\caption{\label{fig5}Neutron diffraction patterns of Cd(CN)$_2$ at specific key $p,T$ values. (a) Ambient-phase Cd(CN)$_2$-I, indexed according to its $Pn\bar3m$ space-group symmetry. The two reflections marked by an asterisk are the strongest impurity peaks attributed to a small remnant Hg(CN)$_2$ contaminant and are present in all patterns. (b) The low-temperature diffraction pattern obtained under minimum applied pressure. The tick marks denote expected reflection conditions for the $C2/c$ structure discussed in the text, and the red line shows the fit obtained using constrained Rietveld refinement. (c) Ambient-temperature diffraction pattern of Cd(CN)$_2$-II at $p=0.2$\,GPa. The general similarity between this pattern and that in (b) links the low-temperature and mid-pressure regimes in the phase diagram shown in Fig.~\ref{fig4}. (d) Ambient-temperature diffraction pattern of Cd(CN)$_2$-III at $p=0.3$\,GPa.}
\end{figure}

One surprise in this study was the observation of a different low-temperature phase transition at $p=0.005$\,GPa, $T\sim120$\,K to that reported using ambient-pressure neutron diffraction measurements in Ref.~\citenum{Coates_2021}. We contrast in Fig.~\ref{fig5}(a,b) the neutron diffraction patterns measured at a pressure of 0.005\,GPa at 300 and 100\,K. The former can be fit using the conventional $Pn\bar3m$ structure model but the latter is noticeably different from the $I4_1/amd$ neutron diffraction pattern reported  in Ref.~\citenum{Coates_2021}. Instead we found that the low-temperature 0.005\,GPa phase gave a similar diffraction pattern to the high-pressure Cd(CN)$_2$-II phase studied at room temperature (using both HRPD and OSIRIS instruments) [Fig.~\ref{fig5}(c)]. Hence the Cd(CN)$_2$-II structure appears to be favoured at both modest pressures and low temperatures, but clearly requires some small pressure to be stabilised relative to the $I4_1/amd$ ambient-pressure/low-temperature phase.

Our data were not of sufficient quality to allow structure solution of the Cd(CN)$_2$-II phase. We obtained a reasonable (but by no means perfect) match to the 100\,K / 0.005\,GPa HRPD diffraction pattern using a highly constrained Rietveld refinement. Our approach was as follows. Noting the splitting of the $(202)$ and $(220)$ reflections ($d\sim3.6,3.2$\,\AA, respectively) was consistent with symmetry-lowering of the $I4_1/amd$ cell to a monoclinic structure, we carried out a search of appropriate monoclinic subgroups. Using ISODISTORT \cite{Campbell_2006}, we found that activation of the $\Gamma_4^+$ and $\Gamma_5^+$ strain modes gave a $C2/c$ solution with appropriate peak splitting. The relationship between this $C2/c$ cell and its $I4_1/amd$ parent was as follows:
\begin{equation}
\left[\begin{array}{l}\mathbf a\\ \mathbf b\\ \mathbf c\end{array}\right]_{C2/c}=\left[\begin{array}{ccc}0&0&-1\\-1&-1&0\\-\frac{1}{2}&\frac{1}{2}&\frac{1}{2}\end{array}\right]\left[\begin{array}{l}\mathbf a\\ \mathbf b\\ \mathbf c\end{array}\right]_{I4_1/amd}.
\end{equation}
We then carried out a constrained Rietveld refinement in which the amplitudes of the $\Gamma_4^+$ and $\Gamma_5^+$ strain modes were allowed to vary, together with an overall scale factor and the conventional strain broadening terms. The background was fixed to that refined against the 150\,K data set (\emph{i.e.}\ when Cd(CN)$_2$ is in its ambient-pressure $Pn\bar3m$ structure). Atom coordinates and atomic displacement parameters were constrained to those determined in the 100\,K $I4_1/amd$ refinement of Ref.~\citenum{Coates_2021}. The resulting fit is shown in Fig.~\ref{fig5}(b). Taking into account the deficiencies of this fit, and the (understandably) reduced signal-to-noise of low-temperature high-pressure measurements, our assignment of $C2/c$ as the crystal symmetry of Cd(CN)$_2$-II is only tentative. {\color{black} We note that the peak splitting observed in our HRPD experiments is consistent with that seen at the highest pressures of our OSIRIS measurementst [Fig.~\ref{fig2}(a)].}

Whatever the true structure of Cd(CN)$_2$-II might be, the slope of the I/II phase boundary is nonetheless consistent with the positive value of $(\frac{\partial B}{\partial T})_p$ determined above for Cd(CN)$_2$-I. Our measured value of 0.030(3)\,GPa\,K$^{-1}$ implies a reduction of the bulk modulus of Cd(CN)$_2$-II by about 6\,GPa on cooling from 300 to 100\,K.

For all temperatures studied, we observed a further phase transition to another high-pressure phase at pressures $p_{{\rm c}2}\simeq0.25$\,GPa. This higher-pressure phase, which we denote Cd(CN)$_2$-III, gave a broad diffraction profile that we were unable to index with any confidence [Fig.~\ref{fig5}(d)].

\subsection{DFT calculations}

Quantum chemical calculations were carried out to investigate the effect of ice-like cyanide ordering on the elastic properties of Cd(CN)$_2$. Previous such calculations have been performed using the same high-symmetry cyanide-ordered structure ($P\bar43m$) in which each Cd is coordinated by either four C atoms or four N atoms \cite{Zwanziger_2007,Ding_2008}. The bulk modulus determined in this way was either 33.03 or 41.8\,GPa, depending on whether GGA and LDA functionals were employed \cite{Ding_2008}.

Our understanding of ambient-temperature Cd(CN)$_2$-I is that, despite the presence of cyanide orientational disorder, Cd coordination environments are nonetheless dominated by the CdC$_2$N$_2$ configuration, and so we interrogated the elastic properties of {\color{black}four} different ordered Cd(CN)$_2$ arrangements with this local arrangement. The key results of our calculations are given in Table~\ref{table1}{\color{black}; and the corresponding geometry-optimised structures are described in Appendix~\ref{app}}. All four of the CdC$_2$N$_2$ states we studied gave bulk moduli smaller than that of the $P\bar43m$ structure, even if they remained a little larger than the experimental value of 13.6\,GPa. The key result here is that the adoption of ice-like coordination environments results in mechanical softening of the Cd(CN)$_2$ structure. This observation is consistent with the {\color{black} thermal-hardening effect noted experimentally (albeit {\it via} a mechanism distinct from that proposed in Refs.~\citenum{Morelock_2013,Dove_2016}):} as temperature increases, the relative fraction of ice-rule violations also increases, which (by our calculations) should result in a larger bulk modulus.

We also include in Table~\ref{table1} the coefficients of thermal expansion determined using a quasiharmonic approach; in the same way that DFT overestimates $B$, it therefore also underestimates the magnitude of $\alpha_V$. We expect that the discrepancy between experiment and calculation reflects the importance of anharmonicity in Cd(CN)$_2$, and that finite-temperature simulations (\emph{e.g.}\ molecular dynamics, MD) should provide better agreement between the two. This same effect was noted for Zn(CN)$_2$, where a $\sim$twofold difference in $B_0$ values was found between 0\,K DFT calculations and finite-temperature MD simulations governed by interatomic potentials fitted against those same calculations \cite{Fang_2013}.

\begin{table}
\center
  \caption{Bulk modulus and thermal expansion data for various Cd(CN)$_2$ configurations obtained {\it via} lattice dynamical calculations.}
  \label{table1}
{\color{black}
 \begin{tabular}{lcccc}
   \hline
Configuration & Space group & $B_0$ (GPa) & $B^{\prime}$&  $\alpha_V$ (MK$^{-1}$) \\ \hline

CdC$_2$N$_2$ (1) &$P4_22_12$& 22.2 & $-34$ &$-49$\\
CdC$_2$N$_2$ (2) &$Pbca$& 24.4 & $-31$ &$-46$\\
CdC$_2$N$_2$ (3) &$Pbcn$& 26.1 & $-29$ &$-44$\\
CdC$_2$N$_2$ (4) &$I4_1/amd$& 31.6 & $-21$ &$-34$\\
CdC$_4$/CdN$_4$ &$P\bar43m$&  33.2 & $-20$&$-31$ \\
    \hline \hline
  \end{tabular}
}\end{table}

\subsection{Role of spin-ice physics}

We turn now to couch the pressure-dependent behaviour explicitly in terms of the spin-ice-like Hamiltonian of Equation~\ref{hamil} developed in Ref.~\citenum{Coates_2021}. In that particular study, we treated the electrostatic dipoles of CN$^-$ ion as direct analogues of the magnetic dipoles in dipolar spin-ices \cite{Champion_2002}. QC calculations were used to derive the key interaction parameters $J$, $D$ and $\rm{\Delta}$, from which the resulting thermal phase behaviour was determined. The key feature of the Hamiltonian, parameterised in this way, is that it exhibits a thermal phase transition between the low-temperature $I4_1/amd$ and $Pn\bar3m$ phases. {\color{black}Note that this Hamiltonian does not explicitly consider the secondary dipoles associated with Cd off-centering discussed earlier, albeit that these are implicit in the emergence of two-in-two-out states of CN$^{-}$ dipole arrangements around each Cd centre.}

Our particular interest in the context of the variable-pressure behaviour of Cd(CN)$_2$ is the parameter $\rm{\Delta}$, which quantifies the energy barrier to cyanide reorientation. It is obtained from nudged elastic band calculations, which also provide quantum-chemical evidence for the rotovibrational coupling observed in the correspondence between cyanide ordering and the evolution of the spontaneous strain \cite{Coates_2021}. The Cd$\ldots$Cd distance is reduced during cyanide reorientation and couples to the unit-cell parameter in a simple geometric manner:
\begin{equation}\label{cella}
a = \frac{2}{\sqrt{3}}\left[d^\dag + \langle S_{\parallel}^2\rangle^{\frac{1}{2}}( d_0- d^{\dag})\right].
\end{equation}
Here, $S_\parallel$ is the projection of the cyanide ion pseudovector onto the corresponding Cd$\ldots$Cd axis, which equals 0 at the mid-point of a cyanide-ion flip and 1 when the Cd--C--N--Cd linkage is perfectly linear. {\color{black} By way of context, Monte Carlo simulations driven by the interaction parameters deduced in Ref.~\citenum{Coates_2021} give values of $\langle S_{\parallel}^2\rangle$ of 0.961 and 0.975 at 300 and 200\,K, respectively.} The paramters $d_0$ and $d^\dagger$ correspond to the Cd$\ldots$Cd distances for linear and mid-flip cyanide-ion geometries, respectively {\color{black}(= 5.545 and 4.898\,\AA, respectively \cite{Coates_2021})}. Because cyanide-ion flips act to reduce $a$, they might be expected to occur increasingly often under the application of pressure. 

We can estimate the contribution of cyanide reorientation to the elastic properties explicitly, starting from the thermodynamic relationship between the bulk modulus and the free energy $\phi(V)$ \cite{Dove_2016}:
\begin{equation}
B=V\frac{\partial^2\phi}{\partial^2V}.
\end{equation}
In our model, the free energy associated with cyanide reorientation is $\phi(V) = 4\mathrm{\Delta}\langle S^2_\parallel\rangle$, where the factor of 4 reflects the number of cyanide pseudospins per Cd(CN)$_2$ unit-cell. The expected volume can be extracted from the Cd$\ldots$Cd distance, according to Equation~\eqref{cella}, resulting in a final expression for the bulk modulus:
\begin{equation}\label{iceb}
B = \frac{4\rm{\Delta}}{(d_0-d^{\dag})^2} \left( \frac{-1}{6a} + \frac{2d^{\dag}}{3\sqrt{3}a^2} \right).
\end{equation}
Using $\mathrm{\Delta}=8\,800$\,K {\color{black}= 1.25 $\times$ 10$^{-19}$\,J} \cite{Coates_2021}, this expression gives $B=24$\,GPa for Cd(CN)$_2$, which is consistent with the value obtained directly from the quantum chemical calculations against which the spin-ice parameters were fitted. Hence the spin-ice description of Cd(CN)$_2$ not only describes its thermal behaviour, but it also does a reasonable job of capturing the elastic behaviour of Cd(CN)$_2$-I. As a final point, we note that the relationship given by Eq.~\eqref{iceb} captures within it the warm-hardening effect observed experimentally. As Cd(CN)$_2$-I is heated, $a$ reduces (\emph{i.e.}\ NTE), which in turn increases $B$.

\section{Concluding Remarks}

In this paper we have characterised the low-temperature/low-pressure phase behaviour of cadmium cyanide. Cd(CN)$_2$ clearly exhibits pressure-dependent behaviour that is typical of an NTE material, in that it undergoes two symmetry lowering phase transitions at modest pressures ($<0.3$\,GPa) and displays pressure-induced softening. The particularly extreme softening, $B_0^{\prime}=-26(3)$, is a hallmark of an incipient elastic instability. The bulk modulus $B_0=13.5(2)$\,GPa is especially low---even amongst other flexible framework materials. The metal-organic framework UiO-66, for example, shows very large NTE with $\alpha_V=-70.2$\,MK$^{-1}$ and has a bulk modulus of 26.5\,GPa \cite{Cliffe_2015,Dissegna_2018}. We find that Cd(CN)$_2$ is substantially softer than Zn(CN)$_2$, and the lower pressures of the observed phase transitions reflect the intrinsic elastic instability of this system.

Whereas the thermodynamic behaviour of Zn(CN)$_2$ was shown to be essentially insensitive to cyanide order \cite{Fang_2013}, a key result here is that Cd(CN)$_2$ models with different cyanide-ion orientations give rise to substantially different elastic properties. This finding suggests a strong coupling between cyanide-ion orientations and dynamic phenomena, that is reminiscent of rotovibrational coupling observed in alkali cyanides \cite{LyndenBell_1994}. Potassium cyanide, for example, shows a rich phase behaviour as a function of temperature and pressure owing to competing quadrupolar and dipolar interactions between cyanide ions, with at least seven phases between $0<T<400$\,K and $0<p<3$\,GPa \cite{Bijvoet_1940,Decker_1974,Rowe_1977,Dultz_1978,Stokes_1993}.  A particular challenge in understanding the physics of Cd(CN)$_2$ then lies in the interplay of excitations occurring at vastly different timescales---\emph{i.e.}\ the cyanide-ion flips and phonon-driven fluctuations. This is a point to which we hope to return in future studies.

Here we have intentionally used gas cells to apply hydrostatic pressure to Cd(CN)$_2$. Our motivation in doing so was to avoid the potential complication of interaction between the pressure-transmitting medium (PTM) and Cd(CN)$_2$. In the case of Zn(CN)$_2$, a large number of different high-pressure transformations occur as the PTM is varied \cite{Lapidus_2013}. One expects even more complex behaviour in the case of Cd(CN)$_2$, since the Cd--C/N interactions are weaker than Zn--C/N, the Cd(CN)$_2$ structure is larger and hence more open, and there is a remarkable empirical propensity for Cd(CN)$_2$ to form clathrate structures \cite{Kitazawa_1992,Kitazawa_1994b,Kitazawa_1994c,Kitazawa_1994d,Kitazawa_1995,Iwamoto_1996,Phillips_2008}. Indeed, we cannot rule out the possibility that Cd(CN)$_2$-III phase involves structural reorganisation to include argon. The close relationship between the (presumed) $C2/c$ Cd(CN)$_2$-II structure and the low-temperature $I4_1/amd$ phase suggests that no chemical change has taken place at the I/II transition. A limitation of the use of gas-cell pressure chambers is that we have accessed only a relatively modest pressure range (albeit covering surprisingly rich structural behaviour). Future studies at higher pressures will likely reveal further phase transitions; in particular we anticipate the value of structural studies of the high-pressure phase transitions studied spectroscopically in Ref.~\citenum{Dehnicke_1974}.

\begin{acknowledgments}
The authors gratefully acknowledge financial support from the E.R.C. (Advanced Grant 788144 to A.L.G.), the UKRI through Frontier Research Grant EP/Z534031/1,  the Royal Society through the Faraday Discovery Fellowships Fund, provided by DSIT, and the Leverhulme Trust (Grants RPG-2015-292 and RPG- 2018-268). A.L.G. thanks Hanna Bostr{\"o}m (Stockholm), Nicholas Funnell (ISIS), Elinor Spencer (Virginia Tech), Nancy Ross (Virginia Tech), Andrew Cairns (Imperial), Andrew Seel (UCL) and Arkadiy Simonov (ETH Zurich) for useful discussions. We thank ISIS for the provision of neutron beamtime and the ISIS Pressure Group for technical assistance. The relevant experiment numbers and data locations are as follows. HRPD: RB1720214 (doi:10.5286/ISIS.E.89472889). OSIRIS: RB1820009 (doi:10.5286/ISIS.E.99689788). MARI: RB1920076 (doi:10.5286/ISIS.E.RB1920076) and RB2010632 (doi:10.5286/ISIS.E.RB2010632).

\end{acknowledgments}

\appendix
{\color{black}
\section{Geometry-optimised Cd(CN)$_2$ structures}\label{app}

Crystallographic details of the four Cd(CN)$_2$ structures detailed in Table~\ref{table1} are given below. In each case the corresponding structure contains Cd centres with uniform CdC$_2$N$_2$ coordination.

Structure 1 adopts $P4_22_12$ space-group symmetry with lattice parameters $a=9.086$\,\AA\ and $c=6.565$\,\AA. The Cd position is on Wyckoff site $4e$ with coordinates $(0.767,0.767,0)$; the C and N positions are on Wyckoff sites $8g$ with coordinates $(-0.752,-0.005,0.791)$ and $(0.558,0.753,0.184)$, respectively.

Structure 2 adopts $Pbca$ space-group symmetry with lattice parameters $a=9.071$\,\AA, $b=9.071$\,\AA, and $c=13.172$\,\AA. The approximate coincidence of $a$ and $b$ is fortuitous. All atoms occupy general positions in the unit-cell. The Cd atom coordinates are $(0.517,0.765,0.374)$; the two crystallographically-distinct C-atom sites have coordinates $(0.502,-0.047,0.270)$ and $(0.705,0.751,0.479)$; and the two crystallographically-distinct N-atom sites are at $(0.308,0.753,0.467)$ and $(0.503,0.555,0.284)$.

Structure 3 adopts $Pbcn$ space-group symmetry with lattice parameters $a=9.103$\,\AA, $b=13.065$\,\AA, and $c=9.103$\,\AA. Again the similarity of $a$ and $c$ is fortuitous and not imposed by symmetry. All atoms occupy general positions in the cell. The Cd atom coordinates are $(0.753,0.126,0.474)$. The two C positions are $(0.741,0.024,0.282)$ and $(-0.055,0.227,0.497)$; and the two N positions are $(0.548,0.220,0.493)$ and $(0.739,0.029,0.678)$.

Structure 4 adopts $I4_1/amd$ space-group symmetry with lattice parameters $a=9.231$\,\AA\ and $c=12.740$\,\AA. The Cd atom position occupies the $8e$ Wyckoff site with coordinates $(0,\frac{1}{4},0.625)$. The C and N atoms are both on  $16h$ sites and have coordinates $(0,-0.046,0.786)$ and $(0,0.560,0.265)$, respectively.

In all cases we made use of the {\sc findsym} package to determine crystal structure symmetries \cite{Stokes_2005}.
}

\bibliography{prm_2026_cdcn2vp}

\begin{thebibliography}{86}
\expandafter\ifx\csname natexlab\endcsname\relax\def\natexlab#1{#1}\fi
\expandafter\ifx\csname bibnamefont\endcsname\relax
  \def\bibnamefont#1{#1}\fi
\expandafter\ifx\csname bibfnamefont\endcsname\relax
  \def\bibfnamefont#1{#1}\fi
\expandafter\ifx\csname citenamefont\endcsname\relax
  \def\citenamefont#1{#1}\fi
\expandafter\ifx\csname url\endcsname\relax
  \def\url#1{\texttt{#1}}\fi
\expandafter\ifx\csname urlprefix\endcsname\relax\def\urlprefix{URL }\fi
\providecommand{\bibinfo}[2]{#2}
\providecommand{\eprint}[2][]{\url{#2}}

\bibitem[{\citenamefont{Sleight}(1998)}]{Sleight_1998}
\bibinfo{author}{\bibfnamefont{A.~W.} \bibnamefont{Sleight}},
  \bibinfo{journal}{Ann. Rev. Mater. Sci.} \textbf{\bibinfo{volume}{28}},
  \bibinfo{pages}{29} (\bibinfo{year}{1998}).

\bibitem[{\citenamefont{Evans}(1999)}]{Evans_1999}
\bibinfo{author}{\bibfnamefont{J.~S.~O.} \bibnamefont{Evans}},
  \bibinfo{journal}{J. Chem. Soc., Dalton Trans.}
  \textbf{\bibinfo{volume}{19}}, \bibinfo{pages}{3317} (\bibinfo{year}{1999}).

\bibitem[{\citenamefont{Barrera et~al.}(2005)\citenamefont{Barrera, Bruno,
  Barron, and Allan}}]{Barrera_2005}
\bibinfo{author}{\bibfnamefont{G.~D.} \bibnamefont{Barrera}},
  \bibinfo{author}{\bibfnamefont{J.~A.~O.} \bibnamefont{Bruno}},
  \bibinfo{author}{\bibfnamefont{T.~H.~K.} \bibnamefont{Barron}},
  \bibnamefont{and} \bibinfo{author}{\bibfnamefont{N.~L.} \bibnamefont{Allan}},
  \bibinfo{journal}{J. Phys.: Condens. Matter} \textbf{\bibinfo{volume}{17}},
  \bibinfo{pages}{R217} (\bibinfo{year}{2005}).

\bibitem[{\citenamefont{Takenaka}(2012)}]{Takenaka_2012}
\bibinfo{author}{\bibfnamefont{K.}~\bibnamefont{Takenaka}},
  \bibinfo{journal}{Sci. Tech. Adv. Mater.} \textbf{\bibinfo{volume}{13}},
  \bibinfo{pages}{013001} (\bibinfo{year}{2012}).

\bibitem[{\citenamefont{Lind}(2012)}]{Lind_2012}
\bibinfo{author}{\bibfnamefont{C.}~\bibnamefont{Lind}},
  \bibinfo{journal}{Materials} \textbf{\bibinfo{volume}{5}},
  \bibinfo{pages}{1125} (\bibinfo{year}{2012}).

\bibitem[{\citenamefont{J. et~al.}(2015)\citenamefont{J., Hu, Deng, and
  Xing}}]{Chen_2015}
\bibinfo{author}{\bibfnamefont{C.}~\bibnamefont{J.}},
  \bibinfo{author}{\bibfnamefont{L.}~\bibnamefont{Hu}},
  \bibinfo{author}{\bibfnamefont{J.~X.} \bibnamefont{Deng}}, \bibnamefont{and}
  \bibinfo{author}{\bibfnamefont{X.~R.} \bibnamefont{Xing}},
  \bibinfo{journal}{Chem. Soc. Rev.} \textbf{\bibinfo{volume}{44}},
  \bibinfo{pages}{3522} (\bibinfo{year}{2015}).

\bibitem[{\citenamefont{Dove and Fang}(2016)}]{Dove_2016}
\bibinfo{author}{\bibfnamefont{M.~T.} \bibnamefont{Dove}} \bibnamefont{and}
  \bibinfo{author}{\bibfnamefont{H.}~\bibnamefont{Fang}},
  \bibinfo{journal}{Rep. Prog. Phys.} \textbf{\bibinfo{volume}{79}},
  \bibinfo{pages}{066503} (\bibinfo{year}{2016}).

\bibitem[{\citenamefont{Shugam and Zhdanov}(1945)}]{Shugam_1945}
\bibinfo{author}{\bibfnamefont{E.}~\bibnamefont{Shugam}} \bibnamefont{and}
  \bibinfo{author}{\bibfnamefont{H.}~\bibnamefont{Zhdanov}},
  \bibinfo{journal}{Acta Physiochim. URSS} \textbf{\bibinfo{volume}{20}},
  \bibinfo{pages}{247} (\bibinfo{year}{1945}).

\bibitem[{\citenamefont{Goodwin and Kepert}(2005)}]{Goodwin_2005}
\bibinfo{author}{\bibfnamefont{A.~L.} \bibnamefont{Goodwin}} \bibnamefont{and}
  \bibinfo{author}{\bibfnamefont{C.~J.} \bibnamefont{Kepert}},
  \bibinfo{journal}{Phys. Rev. B} \textbf{\bibinfo{volume}{71}},
  \bibinfo{pages}{140301(R)} (\bibinfo{year}{2005}).

\bibitem[{\citenamefont{Fairbank et~al.}(2012)\citenamefont{Fairbank, Thompson,
  Cooper, and Goodwin}}]{Fairbank_2012}
\bibinfo{author}{\bibfnamefont{V.~E.} \bibnamefont{Fairbank}},
  \bibinfo{author}{\bibfnamefont{A.~L.} \bibnamefont{Thompson}},
  \bibinfo{author}{\bibfnamefont{R.~I.} \bibnamefont{Cooper}},
  \bibnamefont{and} \bibinfo{author}{\bibfnamefont{A.~L.}
  \bibnamefont{Goodwin}}, \bibinfo{journal}{Phys. Rev. B}
  \textbf{\bibinfo{volume}{86}}, \bibinfo{pages}{104113}
  (\bibinfo{year}{2012}).

\bibitem[{\citenamefont{Coates et~al.}(2021{\natexlab{a}})\citenamefont{Coates,
  Murray, Bostr{\"o}m, M., and Goodwin}}]{Coates_2021b}
\bibinfo{author}{\bibfnamefont{C.~S.} \bibnamefont{Coates}},
  \bibinfo{author}{\bibfnamefont{C.~A.} \bibnamefont{Murray}},
  \bibinfo{author}{\bibfnamefont{H.~L.~B.} \bibnamefont{Bostr{\"o}m}},
  \bibinfo{author}{\bibfnamefont{R.~E.} \bibnamefont{M.}}, \bibnamefont{and}
  \bibinfo{author}{\bibfnamefont{A.~L.} \bibnamefont{Goodwin}},
  \bibinfo{journal}{Mater. Horiz.} \textbf{\bibinfo{volume}{8}},
  \bibinfo{pages}{1446} (\bibinfo{year}{2021}{\natexlab{a}}).

\bibitem[{\citenamefont{Mary et~al.}(1996)\citenamefont{Mary, Evans, Vogt, and
  Sleight}}]{Mary_1996}
\bibinfo{author}{\bibfnamefont{T.~A.} \bibnamefont{Mary}},
  \bibinfo{author}{\bibfnamefont{J.~S.~O.} \bibnamefont{Evans}},
  \bibinfo{author}{\bibfnamefont{T.}~\bibnamefont{Vogt}}, \bibnamefont{and}
  \bibinfo{author}{\bibfnamefont{A.~W.} \bibnamefont{Sleight}},
  \bibinfo{journal}{Science} \textbf{\bibinfo{volume}{272}},
  \bibinfo{pages}{90} (\bibinfo{year}{1996}).

\bibitem[{\citenamefont{Greve et~al.}(2010)\citenamefont{Greve, Martin, Lee,
  Chupas, Chapman, and Wilkinson}}]{Greve_2010}
\bibinfo{author}{\bibfnamefont{B.~K.} \bibnamefont{Greve}},
  \bibinfo{author}{\bibfnamefont{K.~L.} \bibnamefont{Martin}},
  \bibinfo{author}{\bibfnamefont{P.~L.} \bibnamefont{Lee}},
  \bibinfo{author}{\bibfnamefont{P.~J.} \bibnamefont{Chupas}},
  \bibinfo{author}{\bibfnamefont{K.~W.} \bibnamefont{Chapman}},
  \bibnamefont{and} \bibinfo{author}{\bibfnamefont{A.~P.}
  \bibnamefont{Wilkinson}}, \bibinfo{journal}{J. Am. Chem. Soc.}
  \textbf{\bibinfo{volume}{132}}, \bibinfo{pages}{15496}
  (\bibinfo{year}{2010}).

\bibitem[{\citenamefont{Coates and Goodwin}(2019)}]{Coates_2019}
\bibinfo{author}{\bibfnamefont{C.~S.} \bibnamefont{Coates}} \bibnamefont{and}
  \bibinfo{author}{\bibfnamefont{A.~L.} \bibnamefont{Goodwin}},
  \bibinfo{journal}{Mater. Horiz.} \textbf{\bibinfo{volume}{6}},
  \bibinfo{pages}{211} (\bibinfo{year}{2019}).

\bibitem[{\citenamefont{Takenaka}(2024)}]{Takenaka_2024}
\bibinfo{author}{\bibfnamefont{K.}~\bibnamefont{Takenaka}},
  \bibinfo{journal}{Mater. Trans.} \textbf{\bibinfo{volume}{65}},
  \bibinfo{pages}{243} (\bibinfo{year}{2024}).

\bibitem[{\citenamefont{Zwanziger}(2007)}]{Zwanziger_2007}
\bibinfo{author}{\bibfnamefont{J.~W.} \bibnamefont{Zwanziger}},
  \bibinfo{journal}{Phys. Rev. B} \textbf{\bibinfo{volume}{76}},
  \bibinfo{pages}{052102} (\bibinfo{year}{2007}).

\bibitem[{\citenamefont{Ding et~al.}(2008)\citenamefont{Ding, Liang, Jia, and
  Du}}]{Ding_2008}
\bibinfo{author}{\bibfnamefont{P.}~\bibnamefont{Ding}},
  \bibinfo{author}{\bibfnamefont{E.~J.} \bibnamefont{Liang}},
  \bibinfo{author}{\bibfnamefont{Y.}~\bibnamefont{Jia}}, \bibnamefont{and}
  \bibinfo{author}{\bibfnamefont{Z.~Y.} \bibnamefont{Du}}, \bibinfo{journal}{J.
  Phys.: Condens. Matter} \textbf{\bibinfo{volume}{20}},
  \bibinfo{pages}{275224} (\bibinfo{year}{2008}).

\bibitem[{\citenamefont{Coates et~al.}(2021{\natexlab{b}})\citenamefont{Coates,
  Baise, Schmutzler, Simonov, Makepeace, Seel, Smith, Playford, Keen, Siegel
  et~al.}}]{Coates_2021}
\bibinfo{author}{\bibfnamefont{C.~S.} \bibnamefont{Coates}},
  \bibinfo{author}{\bibfnamefont{M.}~\bibnamefont{Baise}},
  \bibinfo{author}{\bibfnamefont{A.}~\bibnamefont{Schmutzler}},
  \bibinfo{author}{\bibfnamefont{A.}~\bibnamefont{Simonov}},
  \bibinfo{author}{\bibfnamefont{J.~W.} \bibnamefont{Makepeace}},
  \bibinfo{author}{\bibfnamefont{A.~G.} \bibnamefont{Seel}},
  \bibinfo{author}{\bibfnamefont{R.~I.} \bibnamefont{Smith}},
  \bibinfo{author}{\bibfnamefont{H.~Y.} \bibnamefont{Playford}},
  \bibinfo{author}{\bibfnamefont{D.~A.} \bibnamefont{Keen}},
  \bibinfo{author}{\bibfnamefont{R.}~\bibnamefont{Siegel}},
  \bibnamefont{et~al.}, \bibinfo{journal}{Nat. Commun.}
  \textbf{\bibinfo{volume}{12}}, \bibinfo{pages}{2272}
  (\bibinfo{year}{2021}{\natexlab{b}}).

\bibitem[{\citenamefont{Bramwell and Harris}(1998)}]{Bramwell_1998}
\bibinfo{author}{\bibfnamefont{S.~T.} \bibnamefont{Bramwell}} \bibnamefont{and}
  \bibinfo{author}{\bibfnamefont{M.~J.} \bibnamefont{Harris}},
  \bibinfo{journal}{J. Phys.: Cond. Matt.} \textbf{\bibinfo{volume}{10}},
  \bibinfo{pages}{L215} (\bibinfo{year}{1998}).

\bibitem[{\citenamefont{den Hertog and Gingras}(2000)}]{denHertog_2000}
\bibinfo{author}{\bibfnamefont{B.~C.} \bibnamefont{den Hertog}}
  \bibnamefont{and} \bibinfo{author}{\bibfnamefont{M.~J.~P.}
  \bibnamefont{Gingras}}, \bibinfo{journal}{Phys. Rev. Lett.}
  \textbf{\bibinfo{volume}{84}}, \bibinfo{pages}{3430} (\bibinfo{year}{2000}).

\bibitem[{\citenamefont{Fennell et~al.}(2009)\citenamefont{Fennell, Deen,
  Wildes, Schamlzl, Prabhakaran, Boothroyd, Aldus, McMorrow, and
  Bramwell}}]{Fennell_2009}
\bibinfo{author}{\bibfnamefont{T.}~\bibnamefont{Fennell}},
  \bibinfo{author}{\bibfnamefont{P.~P.} \bibnamefont{Deen}},
  \bibinfo{author}{\bibfnamefont{A.~R.} \bibnamefont{Wildes}},
  \bibinfo{author}{\bibfnamefont{K.}~\bibnamefont{Schamlzl}},
  \bibinfo{author}{\bibfnamefont{D.}~\bibnamefont{Prabhakaran}},
  \bibinfo{author}{\bibfnamefont{A.~T.} \bibnamefont{Boothroyd}},
  \bibinfo{author}{\bibfnamefont{R.~J.} \bibnamefont{Aldus}},
  \bibinfo{author}{\bibfnamefont{D.~F.} \bibnamefont{McMorrow}},
  \bibnamefont{and} \bibinfo{author}{\bibfnamefont{S.~T.}
  \bibnamefont{Bramwell}}, \bibinfo{journal}{Science}
  \textbf{\bibinfo{volume}{326}}, \bibinfo{pages}{415} (\bibinfo{year}{2009}).

\bibitem[{\citenamefont{Tomasello et~al.}(2015)\citenamefont{Tomasello,
  Castelnovo, Moessner, and Quintanilla}}]{Tomasello_2015}
\bibinfo{author}{\bibfnamefont{B.}~\bibnamefont{Tomasello}},
  \bibinfo{author}{\bibfnamefont{C.}~\bibnamefont{Castelnovo}},
  \bibinfo{author}{\bibfnamefont{R.}~\bibnamefont{Moessner}}, \bibnamefont{and}
  \bibinfo{author}{\bibfnamefont{J.}~\bibnamefont{Quintanilla}},
  \bibinfo{journal}{Phys. Rev. B} \textbf{\bibinfo{volume}{92}},
  \bibinfo{pages}{155120} (\bibinfo{year}{2015}).

\bibitem[{\citenamefont{Perottoni and da~Jornada}(1998)}]{Perottoni_1998}
\bibinfo{author}{\bibfnamefont{C.~A.} \bibnamefont{Perottoni}}
  \bibnamefont{and} \bibinfo{author}{\bibfnamefont{J.~A.~H.}
  \bibnamefont{da~Jornada}}, \bibinfo{journal}{Science}
  \textbf{\bibinfo{volume}{280}}, \bibinfo{pages}{886} (\bibinfo{year}{1998}).

\bibitem[{\citenamefont{Jorgensen et~al.}(1999)\citenamefont{Jorgensen, Hu,
  Teslic, Argyriou, Short, Evans, and Sleight}}]{Jorgensen_1999}
\bibinfo{author}{\bibfnamefont{J.~D.} \bibnamefont{Jorgensen}},
  \bibinfo{author}{\bibfnamefont{Z.}~\bibnamefont{Hu}},
  \bibinfo{author}{\bibfnamefont{S.}~\bibnamefont{Teslic}},
  \bibinfo{author}{\bibfnamefont{D.~N.} \bibnamefont{Argyriou}},
  \bibinfo{author}{\bibfnamefont{S.}~\bibnamefont{Short}},
  \bibinfo{author}{\bibfnamefont{J.~S.~O.} \bibnamefont{Evans}},
  \bibnamefont{and} \bibinfo{author}{\bibfnamefont{A.~W.}
  \bibnamefont{Sleight}}, \bibinfo{journal}{Phys. Rev. B}
  \textbf{\bibinfo{volume}{59}}, \bibinfo{pages}{215} (\bibinfo{year}{1999}).

\bibitem[{\citenamefont{Chapman et~al.}(2009)\citenamefont{Chapman, Halder, and
  Chupas}}]{Chapman_2009}
\bibinfo{author}{\bibfnamefont{K.~W.} \bibnamefont{Chapman}},
  \bibinfo{author}{\bibfnamefont{G.~J.} \bibnamefont{Halder}},
  \bibnamefont{and} \bibinfo{author}{\bibfnamefont{P.~J.}
  \bibnamefont{Chupas}}, \bibinfo{journal}{J. Am. Chem. Soc.}
  \textbf{\bibinfo{volume}{131}}, \bibinfo{pages}{17546}
  (\bibinfo{year}{2009}).

\bibitem[{\citenamefont{Poswal et~al.}(2009)\citenamefont{Poswal, Tyagi, Lausi,
  Deb, and Sharma}}]{Poswal_2009}
\bibinfo{author}{\bibfnamefont{H.~K.} \bibnamefont{Poswal}},
  \bibinfo{author}{\bibfnamefont{A.~K.} \bibnamefont{Tyagi}},
  \bibinfo{author}{\bibfnamefont{A.}~\bibnamefont{Lausi}},
  \bibinfo{author}{\bibfnamefont{S.~K.} \bibnamefont{Deb}}, \bibnamefont{and}
  \bibinfo{author}{\bibfnamefont{S.~M.} \bibnamefont{Sharma}},
  \bibinfo{journal}{J. Solid State Chem.} \textbf{\bibinfo{volume}{182}},
  \bibinfo{pages}{136} (\bibinfo{year}{2009}).

\bibitem[{\citenamefont{Fang et~al.}(2013{\natexlab{a}})\citenamefont{Fang,
  Dove, Rimmer, and Misquitta}}]{Fang_2013}
\bibinfo{author}{\bibfnamefont{H.}~\bibnamefont{Fang}},
  \bibinfo{author}{\bibfnamefont{M.~T.} \bibnamefont{Dove}},
  \bibinfo{author}{\bibfnamefont{L.~H.~N.} \bibnamefont{Rimmer}},
  \bibnamefont{and} \bibinfo{author}{\bibfnamefont{A.~J.}
  \bibnamefont{Misquitta}}, \bibinfo{journal}{Phys. Rev. B}
  \textbf{\bibinfo{volume}{88}}, \bibinfo{pages}{104306}
  (\bibinfo{year}{2013}{\natexlab{a}}).

\bibitem[{\citenamefont{Ma et~al.}(2025)\citenamefont{Ma, Baxter, Park,
  Chariton, dos Santos, Molaison, and Wilkinson}}]{Ma_2025}
\bibinfo{author}{\bibfnamefont{S.~Y.} \bibnamefont{Ma}},
  \bibinfo{author}{\bibfnamefont{S.~J.} \bibnamefont{Baxter}},
  \bibinfo{author}{\bibfnamefont{C.}~\bibnamefont{Park}},
  \bibinfo{author}{\bibfnamefont{S.}~\bibnamefont{Chariton}},
  \bibinfo{author}{\bibfnamefont{A.~M.} \bibnamefont{dos Santos}},
  \bibinfo{author}{\bibfnamefont{J.~J.} \bibnamefont{Molaison}},
  \bibnamefont{and} \bibinfo{author}{\bibfnamefont{A.~P.}
  \bibnamefont{Wilkinson}}, \bibinfo{journal}{Chem. Mater.}
  \textbf{\bibinfo{volume}{37}}, \bibinfo{pages}{1657} (\bibinfo{year}{2025}).

\bibitem[{\citenamefont{Mittal et~al.}(2009)\citenamefont{Mittal, Chaplot, and
  Schober}}]{Mittal_2009}
\bibinfo{author}{\bibfnamefont{R.}~\bibnamefont{Mittal}},
  \bibinfo{author}{\bibfnamefont{S.~L.} \bibnamefont{Chaplot}},
  \bibnamefont{and} \bibinfo{author}{\bibfnamefont{H.}~\bibnamefont{Schober}},
  \bibinfo{journal}{Appl. Phys. Lett.} \textbf{\bibinfo{volume}{95}},
  \bibinfo{pages}{201901} (\bibinfo{year}{2009}).

\bibitem[{\citenamefont{Collings et~al.}(2013)\citenamefont{Collings, Cairns,
  Thompson, Parker, Tang, Tucker, Catafesta, Levelut, Haines, Dmitriev
  et~al.}}]{Collings_2013}
\bibinfo{author}{\bibfnamefont{I.~E.} \bibnamefont{Collings}},
  \bibinfo{author}{\bibfnamefont{A.~B.} \bibnamefont{Cairns}},
  \bibinfo{author}{\bibfnamefont{A.~L.} \bibnamefont{Thompson}},
  \bibinfo{author}{\bibfnamefont{J.~E.} \bibnamefont{Parker}},
  \bibinfo{author}{\bibfnamefont{C.~C.} \bibnamefont{Tang}},
  \bibinfo{author}{\bibfnamefont{M.~G.} \bibnamefont{Tucker}},
  \bibinfo{author}{\bibfnamefont{J.}~\bibnamefont{Catafesta}},
  \bibinfo{author}{\bibfnamefont{C.}~\bibnamefont{Levelut}},
  \bibinfo{author}{\bibfnamefont{J.}~\bibnamefont{Haines}},
  \bibinfo{author}{\bibfnamefont{V.}~\bibnamefont{Dmitriev}},
  \bibnamefont{et~al.}, \bibinfo{journal}{J. Am. Chem. Soc.}
  \textbf{\bibinfo{volume}{135}}, \bibinfo{pages}{7610} (\bibinfo{year}{2013}).

\bibitem[{\citenamefont{Secco et~al.}(2001)\citenamefont{Secco, Liu, Imanaka,
  and Adachi}}]{Secco_2001}
\bibinfo{author}{\bibfnamefont{R.~A.} \bibnamefont{Secco}},
  \bibinfo{author}{\bibfnamefont{J.}~\bibnamefont{Liu}},
  \bibinfo{author}{\bibfnamefont{N.}~\bibnamefont{Imanaka}}, \bibnamefont{and}
  \bibinfo{author}{\bibfnamefont{G.}~\bibnamefont{Adachi}},
  \bibinfo{journal}{J. Mater. Sci. Lett.} \textbf{\bibinfo{volume}{20}},
  \bibinfo{pages}{1339} (\bibinfo{year}{2001}).

\bibitem[{\citenamefont{Keen et~al.}(2007)\citenamefont{Keen, Goodwin, Tucker,
  Dove, Evans, Crichton, and Brunelli}}]{Keen_2007}
\bibinfo{author}{\bibfnamefont{D.~A.} \bibnamefont{Keen}},
  \bibinfo{author}{\bibfnamefont{A.~L.} \bibnamefont{Goodwin}},
  \bibinfo{author}{\bibfnamefont{M.~G.} \bibnamefont{Tucker}},
  \bibinfo{author}{\bibfnamefont{M.~T.} \bibnamefont{Dove}},
  \bibinfo{author}{\bibfnamefont{J.~S.~O.} \bibnamefont{Evans}},
  \bibinfo{author}{\bibfnamefont{W.~A.} \bibnamefont{Crichton}},
  \bibnamefont{and} \bibinfo{author}{\bibfnamefont{M.}~\bibnamefont{Brunelli}},
  \bibinfo{journal}{Phys. Rev. Lett.} \textbf{\bibinfo{volume}{98}},
  \bibinfo{pages}{225501} (\bibinfo{year}{2007}).

\bibitem[{\citenamefont{Hu and Zhang}(2010)}]{Hu_2010}
\bibinfo{author}{\bibfnamefont{Y.~H.} \bibnamefont{Hu}} \bibnamefont{and}
  \bibinfo{author}{\bibfnamefont{L.}~\bibnamefont{Zhang}},
  \bibinfo{journal}{Phys. Rev. B} \textbf{\bibinfo{volume}{81}},
  \bibinfo{pages}{174103} (\bibinfo{year}{2010}).

\bibitem[{\citenamefont{Fang and Dove}(2013)}]{Fang_2013c}
\bibinfo{author}{\bibfnamefont{H.}~\bibnamefont{Fang}} \bibnamefont{and}
  \bibinfo{author}{\bibfnamefont{M.~T.} \bibnamefont{Dove}},
  \bibinfo{journal}{Phys. Rev. B} \textbf{\bibinfo{volume}{87}},
  \bibinfo{pages}{214109} (\bibinfo{year}{2013}).

\bibitem[{\citenamefont{Coates et~al.}(2018)\citenamefont{Coates, Makepeace,
  Seel, Baise, Slater, and Goodwin}}]{Coates_2018}
\bibinfo{author}{\bibfnamefont{C.~S.} \bibnamefont{Coates}},
  \bibinfo{author}{\bibfnamefont{J.~W.} \bibnamefont{Makepeace}},
  \bibinfo{author}{\bibfnamefont{A.~G.} \bibnamefont{Seel}},
  \bibinfo{author}{\bibfnamefont{M.}~\bibnamefont{Baise}},
  \bibinfo{author}{\bibfnamefont{B.}~\bibnamefont{Slater}}, \bibnamefont{and}
  \bibinfo{author}{\bibfnamefont{A.~L.} \bibnamefont{Goodwin}},
  \bibinfo{journal}{Dalton Trans.} \textbf{\bibinfo{volume}{47}},
  \bibinfo{pages}{7263} (\bibinfo{year}{2018}).

\bibitem[{\citenamefont{Zhdanov}(1941)}]{Zhdanov_1941}
\bibinfo{author}{\bibfnamefont{H.}~\bibnamefont{Zhdanov}}, \bibinfo{journal}{C.
  R. Acad. Sci. URSS} \textbf{\bibinfo{volume}{31}}, \bibinfo{pages}{352}
  (\bibinfo{year}{1941}).

\bibitem[{\citenamefont{Williams et~al.}(1997)\citenamefont{Williams, Partin,
  Lincoln, Kouvetakis, and O'Keeffe}}]{Williams_1997}
\bibinfo{author}{\bibfnamefont{D.}~\bibnamefont{Williams}},
  \bibinfo{author}{\bibfnamefont{D.~E.} \bibnamefont{Partin}},
  \bibinfo{author}{\bibfnamefont{F.~J.} \bibnamefont{Lincoln}},
  \bibinfo{author}{\bibfnamefont{J.}~\bibnamefont{Kouvetakis}},
  \bibnamefont{and} \bibinfo{author}{\bibfnamefont{M.}~\bibnamefont{O'Keeffe}},
  \bibinfo{journal}{J. Solid State Chem.} \textbf{\bibinfo{volume}{134}},
  \bibinfo{pages}{164} (\bibinfo{year}{1997}).

\bibitem[{\citenamefont{Manson et~al.}(1998)\citenamefont{Manson, Buschmann,
  and Miller}}]{Manson_1998}
\bibinfo{author}{\bibfnamefont{J.~L.} \bibnamefont{Manson}},
  \bibinfo{author}{\bibfnamefont{W.~E.} \bibnamefont{Buschmann}},
  \bibnamefont{and} \bibinfo{author}{\bibfnamefont{J.~S.}
  \bibnamefont{Miller}}, \bibinfo{journal}{Angew. Chem. Int. Ed. Engl.}
  \textbf{\bibinfo{volume}{37}}, \bibinfo{pages}{783} (\bibinfo{year}{1998}).

\bibitem[{\citenamefont{Nishikiori et~al.}(1991)\citenamefont{Nishikiori,
  Ratcliffe, and Ripmeester}}]{Nishikiori_1991}
\bibinfo{author}{\bibfnamefont{S.}~\bibnamefont{Nishikiori}},
  \bibinfo{author}{\bibfnamefont{C.~I.} \bibnamefont{Ratcliffe}},
  \bibnamefont{and} \bibinfo{author}{\bibfnamefont{J.~A.}
  \bibnamefont{Ripmeester}}, \bibinfo{journal}{J. Chem. Soc., Chem. Commun.}
  pp. \bibinfo{pages}{735--736} (\bibinfo{year}{1991}).

\bibitem[{\citenamefont{Hibble et~al.}(2013)\citenamefont{Hibble, Chippindale,
  Marelli, Kroeker, Michaelis, Greer, Aguiar, Bilb{\'e}, Barney, and
  Hannon}}]{Hibble_2013}
\bibinfo{author}{\bibfnamefont{S.~J.} \bibnamefont{Hibble}},
  \bibinfo{author}{\bibfnamefont{A.~M.} \bibnamefont{Chippindale}},
  \bibinfo{author}{\bibfnamefont{E.}~\bibnamefont{Marelli}},
  \bibinfo{author}{\bibfnamefont{S.}~\bibnamefont{Kroeker}},
  \bibinfo{author}{\bibfnamefont{V.~K.} \bibnamefont{Michaelis}},
  \bibinfo{author}{\bibfnamefont{B.~J.} \bibnamefont{Greer}},
  \bibinfo{author}{\bibfnamefont{P.~M.} \bibnamefont{Aguiar}},
  \bibinfo{author}{\bibfnamefont{E.~J.} \bibnamefont{Bilb{\'e}}},
  \bibinfo{author}{\bibfnamefont{E.~R.} \bibnamefont{Barney}},
  \bibnamefont{and} \bibinfo{author}{\bibfnamefont{A.~C.}
  \bibnamefont{Hannon}}, \bibinfo{journal}{J. Am. Chem. Soc.}
  \textbf{\bibinfo{volume}{135}}, \bibinfo{pages}{16478}
  (\bibinfo{year}{2013}).

\bibitem[{\citenamefont{Hancock et~al.}(2015)\citenamefont{Hancock, Chapman,
  Halder, Morelock, Karlan, Gallington, Bongiorno, Han, Zhou, and
  Wilkinson}}]{Hancock_2015}
\bibinfo{author}{\bibfnamefont{J.~C.} \bibnamefont{Hancock}},
  \bibinfo{author}{\bibfnamefont{K.~W.} \bibnamefont{Chapman}},
  \bibinfo{author}{\bibfnamefont{G.~J.} \bibnamefont{Halder}},
  \bibinfo{author}{\bibfnamefont{C.~R.} \bibnamefont{Morelock}},
  \bibinfo{author}{\bibfnamefont{B.~S.} \bibnamefont{Karlan}},
  \bibinfo{author}{\bibfnamefont{L.~C.} \bibnamefont{Gallington}},
  \bibinfo{author}{\bibfnamefont{A.}~\bibnamefont{Bongiorno}},
  \bibinfo{author}{\bibfnamefont{C.}~\bibnamefont{Han}},
  \bibinfo{author}{\bibfnamefont{S.}~\bibnamefont{Zhou}}, \bibnamefont{and}
  \bibinfo{author}{\bibfnamefont{A.~P.} \bibnamefont{Wilkinson}},
  \bibinfo{journal}{Chem. Mater.} \textbf{\bibinfo{volume}{27}},
  \bibinfo{pages}{3912} (\bibinfo{year}{2015}).

\bibitem[{\citenamefont{Gr{\"u}neisen}(1926)}]{Gruneisen_1926}
\bibinfo{author}{\bibfnamefont{E.}~\bibnamefont{Gr{\"u}neisen}}, in
  \emph{\bibinfo{booktitle}{Thermische Eigenschaften der Stoffe}}, edited by
  \bibinfo{editor}{\bibfnamefont{H.}~\bibnamefont{Geiger}} \bibnamefont{and}
  \bibinfo{editor}{\bibfnamefont{K.}~\bibnamefont{Scheel}}
  (\bibinfo{publisher}{Springer}, \bibinfo{address}{Berlin},
  \bibinfo{year}{1926}), vol.~\bibinfo{volume}{X} of
  \emph{\bibinfo{series}{Handbuch der Physik}}, pp. \bibinfo{pages}{1--59}.

\bibitem[{\citenamefont{Goodwin}(2006)}]{Goodwin_2006}
\bibinfo{author}{\bibfnamefont{A.~L.} \bibnamefont{Goodwin}},
  \bibinfo{journal}{Phys. Rev. B} \textbf{\bibinfo{volume}{74}},
  \bibinfo{pages}{134302} (\bibinfo{year}{2006}).

\bibitem[{\citenamefont{Werner-Zwanziger
  et~al.}(2012)\citenamefont{Werner-Zwanziger, Chapman, and
  Zwanziger}}]{WernerZwanziger_2012}
\bibinfo{author}{\bibfnamefont{U.}~\bibnamefont{Werner-Zwanziger}},
  \bibinfo{author}{\bibfnamefont{K.~W.} \bibnamefont{Chapman}},
  \bibnamefont{and} \bibinfo{author}{\bibfnamefont{J.~W.}
  \bibnamefont{Zwanziger}}, \bibinfo{journal}{Z. Phys. Chem.}
  \textbf{\bibinfo{volume}{226}}, \bibinfo{pages}{1205} (\bibinfo{year}{2012}).

\bibitem[{\citenamefont{Kuhs et~al.}(1984)\citenamefont{Kuhs, Finney, Vettier,
  and Bliss}}]{Kuhs_1984}
\bibinfo{author}{\bibfnamefont{W.~F.} \bibnamefont{Kuhs}},
  \bibinfo{author}{\bibfnamefont{J.~L.} \bibnamefont{Finney}},
  \bibinfo{author}{\bibfnamefont{C.}~\bibnamefont{Vettier}}, \bibnamefont{and}
  \bibinfo{author}{\bibfnamefont{D.~V.} \bibnamefont{Bliss}},
  \bibinfo{journal}{J. Chem. Phys.} \textbf{\bibinfo{volume}{81}},
  \bibinfo{pages}{3612} (\bibinfo{year}{1984}).

\bibitem[{\citenamefont{Jorgensen et~al.}(1984)\citenamefont{Jorgensen,
  Beyerlein, Watanabe, and Worlton}}]{Jorgensen_1984}
\bibinfo{author}{\bibfnamefont{J.~D.} \bibnamefont{Jorgensen}},
  \bibinfo{author}{\bibfnamefont{R.~A.} \bibnamefont{Beyerlein}},
  \bibinfo{author}{\bibfnamefont{N.}~\bibnamefont{Watanabe}}, \bibnamefont{and}
  \bibinfo{author}{\bibfnamefont{T.~G.} \bibnamefont{Worlton}},
  \bibinfo{journal}{J. Chem. Phys.} \textbf{\bibinfo{volume}{81}},
  \bibinfo{pages}{3211} (\bibinfo{year}{1984}).

\bibitem[{\citenamefont{Kuo and Klein}(2004)}]{Kuo_2004}
\bibinfo{author}{\bibfnamefont{J.-L.} \bibnamefont{Kuo}} \bibnamefont{and}
  \bibinfo{author}{\bibfnamefont{M.~L.} \bibnamefont{Klein}},
  \bibinfo{journal}{J. Phys. Chem. B} \textbf{\bibinfo{volume}{108}},
  \bibinfo{pages}{19634} (\bibinfo{year}{2004}).

\bibitem[{\citenamefont{Dehnicke et~al.}(1974)\citenamefont{Dehnicke, Dehnicke,
  Ahsbahs, and Hellner}}]{Dehnicke_1974}
\bibinfo{author}{\bibfnamefont{G.}~\bibnamefont{Dehnicke}},
  \bibinfo{author}{\bibfnamefont{K.}~\bibnamefont{Dehnicke}},
  \bibinfo{author}{\bibfnamefont{H.}~\bibnamefont{Ahsbahs}}, \bibnamefont{and}
  \bibinfo{author}{\bibfnamefont{E.}~\bibnamefont{Hellner}},
  \bibinfo{journal}{Ber. Bunsenges. Phys. Chem.} \textbf{\bibinfo{volume}{78}},
  \bibinfo{pages}{1010} (\bibinfo{year}{1974}).

\bibitem[{\citenamefont{Chapman and Chupas}(2007)}]{Chapman_2007}
\bibinfo{author}{\bibfnamefont{K.~W.} \bibnamefont{Chapman}} \bibnamefont{and}
  \bibinfo{author}{\bibfnamefont{P.~J.} \bibnamefont{Chupas}},
  \bibinfo{journal}{J. Am. Chem. Soc.} \textbf{\bibinfo{volume}{129}},
  \bibinfo{pages}{10090} (\bibinfo{year}{2007}).

\bibitem[{\citenamefont{Ravindran et~al.}(2007)\citenamefont{Ravindran, Arora,
  Chandra, Valsakumar, and Shekar}}]{Ravindran_2007b}
\bibinfo{author}{\bibfnamefont{T.~R.} \bibnamefont{Ravindran}},
  \bibinfo{author}{\bibfnamefont{A.~K.} \bibnamefont{Arora}},
  \bibinfo{author}{\bibfnamefont{S.}~\bibnamefont{Chandra}},
  \bibinfo{author}{\bibfnamefont{M.~C.} \bibnamefont{Valsakumar}},
  \bibnamefont{and} \bibinfo{author}{\bibfnamefont{N.~V.~C.}
  \bibnamefont{Shekar}}, \bibinfo{journal}{Phys. Rev. B}
  \textbf{\bibinfo{volume}{76}}, \bibinfo{pages}{054302}
  (\bibinfo{year}{2007}).

\bibitem[{\citenamefont{Mittal et~al.}(2011)\citenamefont{Mittal, Zbiri,
  Schober, Marelli, Hibble, Chippindale, and Chaplot}}]{Mittal_2011}
\bibinfo{author}{\bibfnamefont{R.}~\bibnamefont{Mittal}},
  \bibinfo{author}{\bibfnamefont{M.}~\bibnamefont{Zbiri}},
  \bibinfo{author}{\bibfnamefont{H.}~\bibnamefont{Schober}},
  \bibinfo{author}{\bibfnamefont{E.}~\bibnamefont{Marelli}},
  \bibinfo{author}{\bibfnamefont{S.~J.} \bibnamefont{Hibble}},
  \bibinfo{author}{\bibfnamefont{A.~M.} \bibnamefont{Chippindale}},
  \bibnamefont{and} \bibinfo{author}{\bibfnamefont{S.~L.}
  \bibnamefont{Chaplot}}, \bibinfo{journal}{Phys. Rev. B}
  \textbf{\bibinfo{volume}{83}}, \bibinfo{pages}{024301}
  (\bibinfo{year}{2011}).

\bibitem[{\citenamefont{Lapidus et~al.}(2013)\citenamefont{Lapidus, Halder,
  Chupas, and Chapman}}]{Lapidus_2013}
\bibinfo{author}{\bibfnamefont{S.~H.} \bibnamefont{Lapidus}},
  \bibinfo{author}{\bibfnamefont{G.~J.} \bibnamefont{Halder}},
  \bibinfo{author}{\bibfnamefont{P.~J.} \bibnamefont{Chupas}},
  \bibnamefont{and} \bibinfo{author}{\bibfnamefont{K.~W.}
  \bibnamefont{Chapman}}, \bibinfo{journal}{J. Am. Chem. Soc.}
  \textbf{\bibinfo{volume}{135}}, \bibinfo{pages}{7621} (\bibinfo{year}{2013}).

\bibitem[{\citenamefont{Fang et~al.}(2013{\natexlab{b}})\citenamefont{Fang,
  Phillips, Dove, Tucker, and Goodwin}}]{Fang_2013b}
\bibinfo{author}{\bibfnamefont{H.}~\bibnamefont{Fang}},
  \bibinfo{author}{\bibfnamefont{A.~E.} \bibnamefont{Phillips}},
  \bibinfo{author}{\bibfnamefont{M.~T.} \bibnamefont{Dove}},
  \bibinfo{author}{\bibfnamefont{M.~G.} \bibnamefont{Tucker}},
  \bibnamefont{and} \bibinfo{author}{\bibfnamefont{A.~L.}
  \bibnamefont{Goodwin}}, \bibinfo{journal}{Phys. Rev. B}
  \textbf{\bibinfo{volume}{88}}, \bibinfo{pages}{144103}
  (\bibinfo{year}{2013}{\natexlab{b}}).

\bibitem[{\citenamefont{Halevy et~al.}(2010)\citenamefont{Halevy, Zamir,
  Winterrose, Sanjit, Grandini, and Moreno-Gobbi}}]{Halevy_2010}
\bibinfo{author}{\bibfnamefont{I.}~\bibnamefont{Halevy}},
  \bibinfo{author}{\bibfnamefont{G.}~\bibnamefont{Zamir}},
  \bibinfo{author}{\bibfnamefont{M.}~\bibnamefont{Winterrose}},
  \bibinfo{author}{\bibfnamefont{G.}~\bibnamefont{Sanjit}},
  \bibinfo{author}{\bibfnamefont{C.~R.} \bibnamefont{Grandini}},
  \bibnamefont{and}
  \bibinfo{author}{\bibfnamefont{A.}~\bibnamefont{Moreno-Gobbi}},
  \bibinfo{journal}{J. Phys.: Conf. Series} \textbf{\bibinfo{volume}{215}},
  \bibinfo{pages}{012013} (\bibinfo{year}{2010}).

\bibitem[{\citenamefont{Coelho}(2007)}]{TOPAS}
\bibinfo{author}{\bibfnamefont{A.~A.} \bibnamefont{Coelho}},
  \emph{\bibinfo{title}{{TOPAS-Academic, version 4.1 (Computer Software)}}}
  (\bibinfo{year}{2007}).

\bibitem[{\citenamefont{Bl{\"o}chl}(1994)}]{Blochl_1994}
\bibinfo{author}{\bibfnamefont{P.~E.} \bibnamefont{Bl{\"o}chl}},
  \bibinfo{journal}{Phys. Rev. B} \textbf{\bibinfo{volume}{50}},
  \bibinfo{pages}{17953} (\bibinfo{year}{1994}).

\bibitem[{\citenamefont{Kresse and Hafner}(1993)}]{Kresse_1993}
\bibinfo{author}{\bibfnamefont{G.}~\bibnamefont{Kresse}} \bibnamefont{and}
  \bibinfo{author}{\bibfnamefont{J.}~\bibnamefont{Hafner}},
  \bibinfo{journal}{Phys. Rev. B} \textbf{\bibinfo{volume}{47}},
  \bibinfo{pages}{558} (\bibinfo{year}{1993}).

\bibitem[{\citenamefont{Kresse and
  Furthm{\"u}ller}(1996{\natexlab{a}})}]{Kresse_1996}
\bibinfo{author}{\bibfnamefont{G.}~\bibnamefont{Kresse}} \bibnamefont{and}
  \bibinfo{author}{\bibfnamefont{J.}~\bibnamefont{Furthm{\"u}ller}},
  \bibinfo{journal}{Phys. Rev. B} \textbf{\bibinfo{volume}{54}},
  \bibinfo{pages}{11169} (\bibinfo{year}{1996}{\natexlab{a}}).

\bibitem[{\citenamefont{Kresse and
  Furthm{\"u}ller}(1996{\natexlab{b}})}]{Kresse_1996b}
\bibinfo{author}{\bibfnamefont{G.}~\bibnamefont{Kresse}} \bibnamefont{and}
  \bibinfo{author}{\bibfnamefont{J.}~\bibnamefont{Furthm{\"u}ller}},
  \bibinfo{journal}{Comput. Mater. Sci.} \textbf{\bibinfo{volume}{6}},
  \bibinfo{pages}{15} (\bibinfo{year}{1996}{\natexlab{b}}).

\bibitem[{\citenamefont{Perdew et~al.}(1996)\citenamefont{Perdew, Burke, and
  Ernzerhof}}]{Perdew_1996}
\bibinfo{author}{\bibfnamefont{J.~P.} \bibnamefont{Perdew}},
  \bibinfo{author}{\bibfnamefont{K.}~\bibnamefont{Burke}}, \bibnamefont{and}
  \bibinfo{author}{\bibfnamefont{M.}~\bibnamefont{Ernzerhof}},
  \bibinfo{journal}{Phys. Rev. Lett.} \textbf{\bibinfo{volume}{77}},
  \bibinfo{pages}{3865} (\bibinfo{year}{1996}).

\bibitem[{\citenamefont{Pulay}(1969)}]{Pulay_1969}
\bibinfo{author}{\bibfnamefont{P.}~\bibnamefont{Pulay}}, \bibinfo{journal}{Mol.
  Phys.} \textbf{\bibinfo{volume}{17}}, \bibinfo{pages}{197}
  (\bibinfo{year}{1969}).

\bibitem[{\citenamefont{Hebbache and Zemzemi}(2004)}]{Hebbache_2004}
\bibinfo{author}{\bibfnamefont{M.}~\bibnamefont{Hebbache}} \bibnamefont{and}
  \bibinfo{author}{\bibfnamefont{M.}~\bibnamefont{Zemzemi}},
  \bibinfo{journal}{Phys. Rev. B} \textbf{\bibinfo{volume}{70}},
  \bibinfo{pages}{224107} (\bibinfo{year}{2004}).

\bibitem[{\citenamefont{Togo and Tanaka}(2015)}]{Togo_2015}
\bibinfo{author}{\bibfnamefont{A.}~\bibnamefont{Togo}} \bibnamefont{and}
  \bibinfo{author}{\bibfnamefont{I.}~\bibnamefont{Tanaka}},
  \bibinfo{journal}{Scr. Mater.} \textbf{\bibinfo{volume}{108}},
  \bibinfo{pages}{1} (\bibinfo{year}{2015}).

\bibitem[{\citenamefont{Togo et~al.}(2010)\citenamefont{Togo, Chaput, Tanaka,
  and Hug}}]{Togo_2010}
\bibinfo{author}{\bibfnamefont{A.}~\bibnamefont{Togo}},
  \bibinfo{author}{\bibfnamefont{L.}~\bibnamefont{Chaput}},
  \bibinfo{author}{\bibfnamefont{I.}~\bibnamefont{Tanaka}}, \bibnamefont{and}
  \bibinfo{author}{\bibfnamefont{G.}~\bibnamefont{Hug}},
  \bibinfo{journal}{Phys. Rev. B} \textbf{\bibinfo{volume}{81}},
  \bibinfo{pages}{174301} (\bibinfo{year}{2010}).

\bibitem[{\citenamefont{Birch}(1847)}]{Birch_1947}
\bibinfo{author}{\bibfnamefont{F.}~\bibnamefont{Birch}},
  \bibinfo{journal}{Phys. Rev.} \textbf{\bibinfo{volume}{71}},
  \bibinfo{pages}{809} (\bibinfo{year}{1847}).

\bibitem[{\citenamefont{Sata et~al.}(2002)\citenamefont{Sata, Shen, Rivers, and
  Sutton}}]{Sata_2002}
\bibinfo{author}{\bibfnamefont{N.}~\bibnamefont{Sata}},
  \bibinfo{author}{\bibfnamefont{G.}~\bibnamefont{Shen}},
  \bibinfo{author}{\bibfnamefont{M.~L.} \bibnamefont{Rivers}},
  \bibnamefont{and} \bibinfo{author}{\bibfnamefont{S.~R.}
  \bibnamefont{Sutton}}, \bibinfo{journal}{Phys. Rev. B}
  \textbf{\bibinfo{volume}{65}}, \bibinfo{pages}{104114}
  (\bibinfo{year}{2002}).

\bibitem[{\citenamefont{Morelock et~al.}(2013)\citenamefont{Morelock, Greve,
  Gallington, Chapman, and Wilkinson}}]{Morelock_2013}
\bibinfo{author}{\bibfnamefont{C.~R.} \bibnamefont{Morelock}},
  \bibinfo{author}{\bibfnamefont{B.~K.} \bibnamefont{Greve}},
  \bibinfo{author}{\bibfnamefont{L.~C.} \bibnamefont{Gallington}},
  \bibinfo{author}{\bibfnamefont{K.~W.} \bibnamefont{Chapman}},
  \bibnamefont{and} \bibinfo{author}{\bibfnamefont{A.~P.}
  \bibnamefont{Wilkinson}}, \bibinfo{journal}{J. Appl. Phys.}
  \textbf{\bibinfo{volume}{114}}, \bibinfo{pages}{213501}
  (\bibinfo{year}{2013}).

\bibitem[{\citenamefont{Mittal et~al.}(2001)\citenamefont{Mittal, Chaplot,
  Schober, and Mary}}]{Mittal_2001}
\bibinfo{author}{\bibfnamefont{R.}~\bibnamefont{Mittal}},
  \bibinfo{author}{\bibfnamefont{S.~L.} \bibnamefont{Chaplot}},
  \bibinfo{author}{\bibfnamefont{H.}~\bibnamefont{Schober}}, \bibnamefont{and}
  \bibinfo{author}{\bibfnamefont{T.~A.} \bibnamefont{Mary}},
  \bibinfo{journal}{Phys. Rev. Lett.} \textbf{\bibinfo{volume}{86}},
  \bibinfo{pages}{4692} (\bibinfo{year}{2001}).

\bibitem[{\citenamefont{Cambpell et~al.}(2006)\citenamefont{Cambpell, Stokes,
  Tanner, and Hatch}}]{Campbell_2006}
\bibinfo{author}{\bibfnamefont{B.~J.} \bibnamefont{Cambpell}},
  \bibinfo{author}{\bibfnamefont{H.~T.} \bibnamefont{Stokes}},
  \bibinfo{author}{\bibfnamefont{D.~E.} \bibnamefont{Tanner}},
  \bibnamefont{and} \bibinfo{author}{\bibfnamefont{D.~M.} \bibnamefont{Hatch}},
  \bibinfo{journal}{J. Appl. Cryst.} \textbf{\bibinfo{volume}{39}},
  \bibinfo{pages}{607} (\bibinfo{year}{2006}).

\bibitem[{\citenamefont{Champion et~al.}(2002)\citenamefont{Champion, Bramwell,
  Holdsworth, and Harris}}]{Champion_2002}
\bibinfo{author}{\bibfnamefont{J.~D.~M.} \bibnamefont{Champion}},
  \bibinfo{author}{\bibfnamefont{S.~T.} \bibnamefont{Bramwell}},
  \bibinfo{author}{\bibfnamefont{P.~C.~W.} \bibnamefont{Holdsworth}},
  \bibnamefont{and} \bibinfo{author}{\bibfnamefont{M.~J.}
  \bibnamefont{Harris}}, \bibinfo{journal}{Europhys. Lett.}
  \textbf{\bibinfo{volume}{57}}, \bibinfo{pages}{93} (\bibinfo{year}{2002}).

\bibitem[{\citenamefont{Cliffe et~al.}(2015)\citenamefont{Cliffe, Hill, Murray,
  Coudert, and Goodwin}}]{Cliffe_2015}
\bibinfo{author}{\bibfnamefont{M.~J.} \bibnamefont{Cliffe}},
  \bibinfo{author}{\bibfnamefont{J.~A.} \bibnamefont{Hill}},
  \bibinfo{author}{\bibfnamefont{C.~A.} \bibnamefont{Murray}},
  \bibinfo{author}{\bibfnamefont{F.-X.} \bibnamefont{Coudert}},
  \bibnamefont{and} \bibinfo{author}{\bibfnamefont{A.~L.}
  \bibnamefont{Goodwin}}, \bibinfo{journal}{Phys. Chem. Chem. Phys.}
  \textbf{\bibinfo{volume}{17}}, \bibinfo{pages}{11586} (\bibinfo{year}{2015}).

\bibitem[{\citenamefont{Dissegna et~al.}(2018)\citenamefont{Dissegna,
  Vervoorts, Hobday, D{\"u}ren, Daisenberger, Smith, Fischer, and
  Kieslich}}]{Dissegna_2018}
\bibinfo{author}{\bibfnamefont{S.}~\bibnamefont{Dissegna}},
  \bibinfo{author}{\bibfnamefont{P.}~\bibnamefont{Vervoorts}},
  \bibinfo{author}{\bibfnamefont{C.~L.} \bibnamefont{Hobday}},
  \bibinfo{author}{\bibfnamefont{T.}~\bibnamefont{D{\"u}ren}},
  \bibinfo{author}{\bibfnamefont{D.}~\bibnamefont{Daisenberger}},
  \bibinfo{author}{\bibfnamefont{A.~J.} \bibnamefont{Smith}},
  \bibinfo{author}{\bibfnamefont{R.~A.} \bibnamefont{Fischer}},
  \bibnamefont{and} \bibinfo{author}{\bibfnamefont{G.}~\bibnamefont{Kieslich}},
  \bibinfo{journal}{J. Am. Chem. Soc.} \textbf{\bibinfo{volume}{140}},
  \bibinfo{pages}{11581} (\bibinfo{year}{2018}).

\bibitem[{\citenamefont{Lynden-Bell and Michel}(1994)}]{LyndenBell_1994}
\bibinfo{author}{\bibfnamefont{R.~M.} \bibnamefont{Lynden-Bell}}
  \bibnamefont{and} \bibinfo{author}{\bibfnamefont{K.~H.}
  \bibnamefont{Michel}}, \bibinfo{journal}{Rev. Mod. Phys.}
  \textbf{\bibinfo{volume}{68}}, \bibinfo{pages}{721} (\bibinfo{year}{1994}).

\bibitem[{\citenamefont{Bijvoet and Lely}(1940)}]{Bijvoet_1940}
\bibinfo{author}{\bibfnamefont{J.~M.} \bibnamefont{Bijvoet}} \bibnamefont{and}
  \bibinfo{author}{\bibfnamefont{J.~A.} \bibnamefont{Lely}},
  \bibinfo{journal}{Rec. Trav. Chim. Pays-Bas} \textbf{\bibinfo{volume}{59}},
  \bibinfo{pages}{908} (\bibinfo{year}{1940}).

\bibitem[{\citenamefont{Decker et~al.}(1974)\citenamefont{Decker, Beyerlein,
  Roult, and Worlton}}]{Decker_1974}
\bibinfo{author}{\bibfnamefont{D.~L.} \bibnamefont{Decker}},
  \bibinfo{author}{\bibfnamefont{R.~A.} \bibnamefont{Beyerlein}},
  \bibinfo{author}{\bibfnamefont{G.}~\bibnamefont{Roult}}, \bibnamefont{and}
  \bibinfo{author}{\bibfnamefont{T.~G.} \bibnamefont{Worlton}},
  \bibinfo{journal}{Phys. Rev. B} \textbf{\bibinfo{volume}{10}},
  \bibinfo{pages}{3584} (\bibinfo{year}{1974}).

\bibitem[{\citenamefont{Rowe et~al.}(1977)\citenamefont{Rowe, Rush, and
  Prince}}]{Rowe_1977}
\bibinfo{author}{\bibfnamefont{J.~M.} \bibnamefont{Rowe}},
  \bibinfo{author}{\bibfnamefont{J.~J.} \bibnamefont{Rush}}, \bibnamefont{and}
  \bibinfo{author}{\bibfnamefont{E.}~\bibnamefont{Prince}},
  \bibinfo{journal}{J. Chem. Phys.} \textbf{\bibinfo{volume}{66}},
  \bibinfo{pages}{5147} (\bibinfo{year}{1977}).

\bibitem[{\citenamefont{Dultz and Krause}(1978)}]{Dultz_1978}
\bibinfo{author}{\bibfnamefont{W.}~\bibnamefont{Dultz}} \bibnamefont{and}
  \bibinfo{author}{\bibfnamefont{H.}~\bibnamefont{Krause}},
  \bibinfo{journal}{Phys. Rev. B} \textbf{\bibinfo{volume}{18}},
  \bibinfo{pages}{394} (\bibinfo{year}{1978}).

\bibitem[{\citenamefont{Stokes et~al.}(1993)\citenamefont{Stokes, Decker,
  Nelson, and Jorgensen}}]{Stokes_1993}
\bibinfo{author}{\bibfnamefont{H.~T.} \bibnamefont{Stokes}},
  \bibinfo{author}{\bibfnamefont{D.~L.} \bibnamefont{Decker}},
  \bibinfo{author}{\bibfnamefont{H.~M.} \bibnamefont{Nelson}},
  \bibnamefont{and} \bibinfo{author}{\bibfnamefont{J.~D.}
  \bibnamefont{Jorgensen}}, \bibinfo{journal}{Phys. Rev. B}
  \textbf{\bibinfo{volume}{47}}, \bibinfo{pages}{11082} (\bibinfo{year}{1993}).

\bibitem[{\citenamefont{Kitazawa et~al.}(1992)\citenamefont{Kitazawa,
  Nishikiori, Yamagishi, Kuroda, and Iwamoto}}]{Kitazawa_1992}
\bibinfo{author}{\bibfnamefont{T.}~\bibnamefont{Kitazawa}},
  \bibinfo{author}{\bibfnamefont{S.-I.} \bibnamefont{Nishikiori}},
  \bibinfo{author}{\bibfnamefont{A.}~\bibnamefont{Yamagishi}},
  \bibinfo{author}{\bibfnamefont{R.}~\bibnamefont{Kuroda}}, \bibnamefont{and}
  \bibinfo{author}{\bibfnamefont{T.}~\bibnamefont{Iwamoto}},
  \bibinfo{journal}{J. Chem. Soc., Chem. Commun.} pp. \bibinfo{pages}{413--415}
  (\bibinfo{year}{1992}).

\bibitem[{\citenamefont{Kitazawa
  et~al.}(1994{\natexlab{a}})\citenamefont{Kitazawa, Nishikiori, Kuroda, and
  Iwamoto}}]{Kitazawa_1994b}
\bibinfo{author}{\bibfnamefont{T.}~\bibnamefont{Kitazawa}},
  \bibinfo{author}{\bibfnamefont{S.-I.} \bibnamefont{Nishikiori}},
  \bibinfo{author}{\bibfnamefont{R.}~\bibnamefont{Kuroda}}, \bibnamefont{and}
  \bibinfo{author}{\bibfnamefont{T.}~\bibnamefont{Iwamoto}},
  \bibinfo{journal}{J. Chem. Soc., Dalton Trans.} pp.
  \bibinfo{pages}{1029--1036} (\bibinfo{year}{1994}{\natexlab{a}}).

\bibitem[{\citenamefont{Kitazawa
  et~al.}(1994{\natexlab{b}})\citenamefont{Kitazawa, Kikuyama, Takahashi, and
  Takeda}}]{Kitazawa_1994c}
\bibinfo{author}{\bibfnamefont{T.}~\bibnamefont{Kitazawa}},
  \bibinfo{author}{\bibfnamefont{T.}~\bibnamefont{Kikuyama}},
  \bibinfo{author}{\bibfnamefont{M.}~\bibnamefont{Takahashi}},
  \bibnamefont{and} \bibinfo{author}{\bibfnamefont{M.}~\bibnamefont{Takeda}},
  \bibinfo{journal}{J. Chem. Soc., Dalton Trans.} pp.
  \bibinfo{pages}{2933--2937} (\bibinfo{year}{1994}{\natexlab{b}}).

\bibitem[{\citenamefont{Kitazawa
  et~al.}(1994{\natexlab{c}})\citenamefont{Kitazawa, Nishikiori, and
  Iwamoto}}]{Kitazawa_1994d}
\bibinfo{author}{\bibfnamefont{T.}~\bibnamefont{Kitazawa}},
  \bibinfo{author}{\bibfnamefont{S.-I.} \bibnamefont{Nishikiori}},
  \bibnamefont{and} \bibinfo{author}{\bibfnamefont{T.}~\bibnamefont{Iwamoto}},
  \bibinfo{journal}{J. Chem. Soc., Dalton Trans.} pp.
  \bibinfo{pages}{3695--3710} (\bibinfo{year}{1994}{\natexlab{c}}).

\bibitem[{\citenamefont{Kitazawa et~al.}(1995)\citenamefont{Kitazawa, Kikuyama,
  Takeda, and Iwamoto}}]{Kitazawa_1995}
\bibinfo{author}{\bibfnamefont{T.}~\bibnamefont{Kitazawa}},
  \bibinfo{author}{\bibfnamefont{T.}~\bibnamefont{Kikuyama}},
  \bibinfo{author}{\bibfnamefont{M.}~\bibnamefont{Takeda}}, \bibnamefont{and}
  \bibinfo{author}{\bibfnamefont{T.}~\bibnamefont{Iwamoto}},
  \bibinfo{journal}{J. Chem. Soc., Dalton Trans.} pp.
  \bibinfo{pages}{3715--3720} (\bibinfo{year}{1995}).

\bibitem[{\citenamefont{Iwamoto}(1996)}]{Iwamoto_1996}
\bibinfo{author}{\bibfnamefont{T.}~\bibnamefont{Iwamoto}}, \bibinfo{journal}{J.
  Incl. Phen. Mol. Recog. Chem.} \textbf{\bibinfo{volume}{24}},
  \bibinfo{pages}{61} (\bibinfo{year}{1996}).

\bibitem[{\citenamefont{Phillips et~al.}(2008)\citenamefont{Phillips, Goodwin,
  Halder, Southon, and Kepert}}]{Phillips_2008}
\bibinfo{author}{\bibfnamefont{A.~E.} \bibnamefont{Phillips}},
  \bibinfo{author}{\bibfnamefont{A.~L.} \bibnamefont{Goodwin}},
  \bibinfo{author}{\bibfnamefont{G.~J.} \bibnamefont{Halder}},
  \bibinfo{author}{\bibfnamefont{P.~D.} \bibnamefont{Southon}},
  \bibnamefont{and} \bibinfo{author}{\bibfnamefont{C.~J.}
  \bibnamefont{Kepert}}, \bibinfo{journal}{Angew. Chem. Int. Ed.}
  \textbf{\bibinfo{volume}{47}}, \bibinfo{pages}{1396} (\bibinfo{year}{2008}).

\bibitem[{\citenamefont{Stokes and Hatch}(2005)}]{Stokes_2005}
\bibinfo{author}{\bibfnamefont{H.~T.} \bibnamefont{Stokes}} \bibnamefont{and}
  \bibinfo{author}{\bibfnamefont{D.~M.} \bibnamefont{Hatch}},
  \bibinfo{journal}{J. Appl. Cryst.} \textbf{\bibinfo{volume}{38}},
  \bibinfo{pages}{237} (\bibinfo{year}{2005}).

\end{thebibliography}

\end{document}